\documentclass[10pt,journal,compsoc]{IEEEtran}

\usepackage{booktabs} 

\ifCLASSOPTIONcompsoc
  \usepackage[nocompress]{cite}
\else
  \usepackage{cite}
\fi

\usepackage{amsmath}
\usepackage{comment}
\usepackage{xspace}
\usepackage{listings}
\usepackage{xcolor}
\usepackage{float}
\usepackage{algorithmic}
\usepackage{algorithm}
\usepackage{mdframed}
\usepackage{filecontents}
\usepackage{pgfplots}
\PassOptionsToPackage{hyphens}{url}
\usepackage{url}

\usepackage{array}
\usepackage{multirow}
\usepackage{longtable}
\usepackage{balance}
\usepackage{amssymb}
\usepackage{pifont}

\usepackage{balance}
\usepackage{paralist}
\usepackage{booktabs}
\usepackage{subcaption}

\newcolumntype{P}[1]{>{\centering\arraybackslash}p{#1}}

\definecolor{OliveGreen}{rgb}{0,0.6,0}

\newcommand{\tabincell}[2]{\begin{tabular}{@{}#1@{}}#2\end{tabular}}
\newcommand{\cmark}{\ding{51}}%
\newcommand{\xmark}{\ding{55}}%

\newcommand{\CASE}[1]{\STATE \textbf{case} #1\textbf{:} \begin{ALC@g}}
\newcommand{\ENDCASE}{\end{ALC@g}}

\newcommand{\commentty}[1]{{\color{red} \sf (TY: #1)}}
\newcommand{\wym}[1]{{#1}}
\newcommand{\wymm}[1]{{#1}} 

\newcommand{\subB}{\vspace{3pt}\noindent\textbf}

\newcommand{\ignore}[1]{}
\definecolor{Light}{gray}{.85}

\newcommand{\Name}{SDRacer}

\newcommand{\mylongtitle}[1]{%
	\ifodd\value{page}%
	\protect\parbox{0.97\linewidth}{#1}\hfill%
	\else%
	\hfill\protect\parbox{0.97\linewidth}{#1}%
	\fi%
}

\newcommand{\eg}{\hbox{\emph{e.g.}}\xspace}
\newcommand{\ie}{\hbox{\emph{i.e.}}\xspace}
 
\newcommand{\st}{\hbox{\emph{s.t.}}\xspace}

\newcommand{\etc}{\hbox{\emph{etc.}}\xspace}

\begin{document}
%
\title{Automatic Detection, Validation and Repair of Race Conditions in Interrupt-Driven Embedded Software}

\author{Yu Wang, Fengjuan Gao, Linzhang Wang, Tingting Yu,~\IEEEmembership{Member,~IEEE}, Ke Wang, Jianhua Zhao, and Xuandong Li%
	\thanks{Y. Wang, F. Gao, L. Wang, J. Zhao and X. Li  are with State Key Laboratory of Novel Software Technology, Nanjing~University, China (email:  yuwang\_cs@nju.edu.cn, fjgao@smail.nju.edu.cn, \{lzwang, zhaojh, lxd\}@nju.edu.cn).}%
	\thanks{T. Yu is with University of Kentucky,~USA (email: tyu@cs.uky.edu).}
	\thanks{K. Wang is with Visa Research,~USA (email:  kewang@visa.com).}
}

\markboth{Journal of \LaTeX\ Class Files,~Vol.~14, No.~8, August~2015}%
{Shell \MakeLowercase{\textit{et al.}}: Bare Demo of IEEEtran.cls for Computer Society Journals}
%

\IEEEtitleabstractindextext{%
    \begin{abstract}
	    Interrupt-driven programs are widely deployed in safety-critical embedded systems 
to perform hardware and resource dependent data operation tasks. 
The frequent use of interrupts in these systems can cause race conditions to 
occur due to interactions between application tasks and interrupt handlers \wym{(or two interrupt handlers)}. 
Numerous program analysis and testing techniques have been proposed to 
detect races in multithreaded programs. Little work, however, has addressed race 
condition problems related to hardware interrupts.
In this paper, we present \Name{}, an automated framework that can detect, 
validate and repair race conditions in interrupt-driven embedded software.  
It uses a combination of static analysis and symbolic execution 
to generate input data for exercising the
potential races. It then employs virtual platforms 
to dynamically validate these races by forcing the interrupts to occur
at the potential racing points.
\wym{Finally, it provides repair candidates to eliminate the detected races.
We evaluate \Name{} on nine real-world  embedded programs written in C language.
The results show that \Name{} can precisely detect and successfully fix race conditions.
}
    \end{abstract}
	
	\begin{IEEEkeywords}
		Embedded Software, Interrupts, Race Condition, Software Testing, Repair Suggestion.
\end{IEEEkeywords}}


\maketitle

\IEEEdisplaynontitleabstractindextext

%
\IEEEpeerreviewmaketitle



\IEEEraisesectionheading{\section{Introduction}}
\label{sec:intro}

\IEEEPARstart{M}{odern} embedded systems are highly concurrent, memory, and sensor intensive, and run in resource
constrained environments. They are often programmed using interrupts to provide
concurrency and allow communication with peripheral devices. Typically, a peripheral
device initiates a communication by issuing an interrupt that is then serviced by an
interrupt service routine (ISR), which is a procedure that is invoked when a
particular type of interrupt is issued.  The frequent use of interrupts
can cause concurrency faults such as data races to occur due to interactions
between application tasks and  ISRs.  Such faults are often difficult to detect,
isolate, and correct because they are sensitive to execution interleavings.

As an example, occurrences of race conditions 
between interrupt handlers and applications have been reported in a previous 
release of uCLinux~\cite{JacksonLKML09}, 
a Linux OS designed for real-time embedded systems. 
In this particular case, the serial communication line can be 
shared by an application through a device driver and an interrupt handler. 
In common instances, the execution of both the driver and the 
handler would be correct. However, in an exceptional 
operating scenario, the driver would execute a rarely executed path. 
If an interrupt occurs at that particular time, simultaneous 
transmissions of data is possible (Section~\ref{sec:background} provides
further details).


Many techniques and algorithms have been proposed to 
address concurrency faults, such as race conditions.
These include static analysis~\cite{Praun2003,Flanagan2003,Williams2005b,Naik2009,Joshi2012},
dynamic monitoring~\cite{Kahlon09,Marino2009,Bond2010,Effinger-Dean2012},
schedule exploration~\cite{Visser2003,Sen2007,Musuvathi2008,Sen08,Coons2010,Burckhardt2010a},
test generation~\cite{pldi2012,icse2012-ballerina} and concurrency fault repair~\cite{kelk2013automatically, jin2011automated, liu2012axis, liu2016understanding, surendran2014test}.
These techniques, however, focus on thread-level
races. Applying these directly to interrupt-driven software
is not straightforward.
First, interrupt-driven programs employ a different
concurrency model. The implicit dependencies between asynchronous
concurrency events and their priorities
complicate the happens-before relations that are used for detecting races.
Second, controlling interrupts requires fine-grained execution control; that is,
it must be possible to control execution at the machine code level
and not at the program statement level, which is the granularity at which many 
existing techniques operate.
Third, occurrences of interrupts are highly dependent on hardware states;
that is, interrupts can occur only when hardware components are in certain states.
Existing techniques are often not cognizant of hardware states.
Fourth, repairing race conditions in interrupt-driven embedded systems 
usually requires disabling and enabling interrupt sources in hardware;
this is different from repairing thread-level concurrency faults.

There are several techniques for testing embedded
systems with a particular focus on interrupt-level
concurrency faults~\cite{higashi2010effective, Regehr05, lai2008inter}.
For example,  Higashi et al.~\cite{higashi2010effective} improve
random testing via a mechanism that causes
interrupts to occur at all instruction points to
detect interrupt related data races.
However, these techniques rely on existing test inputs and could miss races
that could otherwise be detected by other inputs. In addition, these techniques
do not account for the implicit dependencies among tasks and interrupts due to
priorities. 
Furthermore, none of the existing techniques can automatically 
repair race conditions.

This paper presents \Name{} (\textbf{S}tatic and \textbf{D}ynamic \textbf{Race} detection), an automated tool that combines static analysis,
symbolic execution, and dynamic simulation to detect, validate and repair race
conditions in interrupt-driven embedded systems.
\Name{} first employs static analysis to identify code locations for potential
races. \Name{} then uses symbolic execution
to generate input data and interrupt interleavings for exercising the potential
racing points; a subset of false positives can be eliminated at this step.
To further validate race warnings, \Name{} leverages
the virtual platform's abilities to interrupt execution without affecting the
states of the virtualized system and to manipulate memory
and buses directly to force interrupts to occur.
Finally, \Name{} provides repair suggestions for each validated race and developers decide which repair strategy is more suitable for the race. 

To evaluate the effectiveness and efficiency of \Name{}, we apply the approach to
nine embedded system benchmarks with 
race conditions.
Our results show that \Name{}  precisely
detected 190 race conditions and successfully
repaired them without causing deadlocks or
excessive performance degradation.
Furthermore, the time taken by \Name{} to detect, validate and repair
 races is typically a few minutes,
indicating that it is efficient enough for practical use.

In summary, this paper contributes the following: \begin{compactitem}
  \item An automated framework that can detect, validate and repair
  race conditions for interrupt-driven embedded software systems\footnote[1]{Partially available at https://github.com/ITWOI/DATA/tree/main/SSD}.


  \item A practical tool for directly handling the C
        code of interrupt-driven embedded software. 
  
  \item Empirical evidence that the approach can
    effectively and efficiently detect and repair race conditions in real-world
    interrupt-driven embedded systems.

\end{compactitem}

The rest of this paper is organized as follows. 
In  the next section
we present a motivating example and background.
We then describe \Name{} in Section \ref{sec:mod}.
Our empirical study follows in Sections \ref{sec:study}
-- \ref{sec:result}, followed by discussion in Section \ref{sec:dis}.
We present related work in Section \ref{sec:ret}, and end with
conclusions in Section \ref{sec:clu}.


\section{Motivation and Background}
\label{sec:background}

In this section we provide background and use an example
to illustrate the challenges
in addressing race conditions in interrupt-driven
embedded software.

\subsection{Interrupt-driven Embedded Systems}
In embedded systems, an interrupt alerts the processor to a high-priority condition requiring
the interruption of the current code the processor is executing.
The processor responds by suspending its current activities, saving its state,
and executing a function called an interrupt handler (or an interrupt service routine, ISR)
to deal with the event. This interruption is temporary, and, after the interrupt handler
finishes, the processor resumes normal activities.

We denote an interrupt-driven program by
$P$ = {\tt Task} $\|$ {\tt ISR},
where {\tt Task} is the main program that consists
of one or more tasks (or threads) and
{\tt ISR} = $ISR_1 \| ISR_2 \| \ldots \| ISR_N$
indicates interrupt service routines. The
subscripts of ISRs indicate interrupt numbers,
with larger numbers denoting lower priorities.
Typically, $P$ receives two types of \emph{incoming data}:
command inputs as entered by users
and sensor inputs such as data received
through specific devices (\eg, a UART port).
An \emph{interrupt schedule} specifies a sequence
of interrupts occurring at specified program locations.
In this work, we do not consider reentrant interrupts
(interrupts that can preempt themselves); these are uncommon
and used only in special situations~\cite{Regehr05}.

\subsection{Race Conditions in Interrupt-driven Programs}
\label{sec:Def}
%



\wymm{
A race condition occurs if two events access a shared resource for which the order of accesses is nondeterministic, i.e., the accesses may happen in either order or simultaneously \cite{vonPraun2011, netzer1992what}. 
It broadly refers to data races (if accessing the shared resource simultaneous) and order violations (if accessing the resources in either order).  
Specifically, in our context, a race condition is reported when two conditions are met: 
1) the execution of
a task or an interrupt handler $T$
is preempted by another interrupt handler $H$ after
a shared memory access $m$, and 2)  $H$
manipulates the content of $m$.
More formally,
}

\begin{center}
\footnotesize
$\mathit{
e_i = MEM (m_i, a_i, T_i, p_i, s_i) \wedge e_j = MEM (m_j,
a_j, T_j, p_j, s_j)}$
$\mathit{\wedge m_i = m_j \wedge  (a_j =
WRITE \vee a_i = WRITE)}$
$\mathit{  \wedge s_i = s_j.enabled \wedge p_j > p_i
}$
\end{center}

%
\noindent
MEM$(m_i, a_i, T_i, p_i, s_i)$ denotes a 
task or an ISR $T_i$ with priority $p_i$ performs
an access $a \in $\{WRITE, READ\} to memory location $m_i$ while
in an hardware state $s_i$. 
The above condition states that two events $e_i$ and $e_j$ are in race condition
if they access the same memory location and at least one access is a write.
Here,
$e_i$ is from a task or an ISR and $e_j$ is from a different ISR,
the interrupt of $H_j$ is enabled when $e_i$
happens, and the priority $p_j$ is greater than $p_i$.


\wymm{
We consider the definition as a variant of order violations.
Data races \cite{pratikakis2006locksmith} are not
applicable between a task and an ISR or between ISRs,
because a memory cannot be simultaneously accessed
by the tasks or the ISRs. That said, a memory is always 
accessed by a task (or a low-priority ISR) and then preempted by an ISR.
Interrupts have an asymmetric preemption relation with the processor's 
non-interrupt context: 
interrupts can preempt non-interrupt activity (\ie, tasks)
but the reverse is not true~\cite{Regehr05}. 
Atomicity violations  \cite{lu2008learning} are not applicable because they require three shared variable accesses.
Traditional order violations \cite{lu2008learning} are also not applicable since there is no enforced execution order.
However, we regard it as a variant of order violations because interrupts have an asymmetric preemption relation with the processor's non-interrupt context.
}

\ignore{A race condition is broadly referred to data races, atomicity violations, and order violations. In this work,
we consider the definition as a variant of order violations.
Data races are not
applicable between a task and an ISR or between ISRs,
because a memory cannot be simultaneously accessed
by the tasks or the ISRs. That said, a memory is always 
accessed by a task (or a low-priority ISR) and then preempted by an ISR.
Interrupts have an asymmetric preemption relation with the processor's 
non-interrupt context: 
interrupts can preempt non-interrupt activity (\ie, tasks)
but the reverse is not true~\cite{Regehr05}. }

\subsection{A Motivating Example}
\label{sec:example}

In prior releases of uCLinux version 2.4, 
there is a particular  race condition that
occurs between the UART driver program uart 
start and the UART ISR serial8250 interrupt \cite{JacksonLKML09}. 
We provide the code snippets
(slightly modified for ease of presentation) 
that illustrate the error in Figure \ref{exp1}. 
The variables marked with bold indicate shared resources
accessed by both tasks and ISRs. 

Under normal operating conditions, 
the interrupt service routines
(ISRs) are always responsible for transmitting data. 
There are two ISRs:
{\tt irq1\_handler} has a higher priority 
than {\tt irq2\_handler}. However, several sources have 
shown that problems such as races with other 
processors on the system or intermittent port 
problems can cause the response from the ISRs to get 
lost or cause a failure to correctly install the ISRs, 
respectively. When that happens, the port is 
registered as ``buggy'' (line 5) and workaround code 
based on polling instead of using interrupts is used (line 12-16). 
Unfortunately, the enabled {\tt irq1\_handler} is not 
disabled in the workaround code region so by the time the 
workaround code is executed, it is possible that 
{\tt irq1\_handler} preempts and modifies the shared
variable \texttt{xmit->tail} (line 14); this causes the serial
port to receive the wrong data (line 15).

\lstset{
  language={[ANSI]C},
  moredelim=**[is][\color{OliveGreen}]{@}{@},
  basicstyle=\fontsize{6}{6}\ttfamily, mathescape,
  breaklines=true,
  showstringspaces=false,
  escapeinside={<@}{@>}
  }
\noindent

\begin{figure}[t]
\small
\begin{mdframed}[roundcorner=5pt]
\scriptsize
\begin{lstlisting}[language=C++,showstringspaces=false]
int transmit(struct uart_port *port){
  ...
  if (<@\textbf{iir}@> & UART_IIR_NO_INT) { 
    if (!(port->bugs & UART_BUG_TXEN)) {
      port->bugs |= UART_BUG_TXEN;
      ...
    }
  }	
  serial_out(port, UART_IER, flags); /*disable irq2*/

  ...
  if (port->bugs & UART_BUG_TXEN) { /*workaround*/
    ... 
    p = <@\textbf{xmit->tail}@> + 1;
    serial_outp(port, UART_TX, p.x_char); /*incorrect output*/ 
  }
}

static irqreturn_t irq1_handler(...){

  if (<@\textbf{thr}@> == 0x1101) {
    <@\textbf{xmit->tail}@> = a + 1;
  }
  b = <@\textbf{xmit->tail}@>;
  ...
}
 
static irqreturn_t irq2_handler(...){ 
    
  if (<@\textbf{thr}@> != 0x1101) {
    <@\textbf{xmit->tail}@> = c + 1;
    ...
  }
}

\end{lstlisting}

\vspace*{-4pt}
\end{mdframed}
\vspace*{-5pt}
\caption{\label{exp1} \textbf{\small Race condition in a UART device driver}}
\normalsize
\end{figure}


The first challenge is that embedded systems use special operations to control
interrupts, some of which may not even be recognized by
existing static and dynamic analysis techniques.
For example, \texttt{serial\_out} 
disables {\tt irq2\_handler} by directly
flagging an interrupt bit at the hardware level using
the variable \texttt{flags} (line 9).
Failing to identify such operations would report
false positives. For example, conservative analysis techniques 
would falsely report that
there is a race condition between line 14 and line 31 on the variable
{\tt xmit->tail} even if the {\tt irq2\_handler} is disabled in the task.
Therefore, hardware states and operations must be known when
testing for race conditions in interrupt-driven embedded systems.

Second,  task and interrupt priorities affect the order relations
between concurrency events. For example,
the content of {\tt xmit->tail} at
line~24 cannot be modified by the write 
of {\tt xmit->tail} at line~31 
due to the reason that the {\tt irq1\_handler} has a higher priority than 
the {\tt irq2\_handler}.
Therefore,
existing techniques that neglect the effect of priorities
would lead to false positives.

Third, exposing this race condition
requires specific input data from the hardware. 
For example, only when the IIR register is
cleared (\ie, \texttt{iir \& UART\_IIR\_NO\_INT}
is true) and the port is set  to
``buggy'' will the true branch (line 4)
be taken in the {\tt transmit} function.
Existing techniques
on testing interrupt-driven programs that
rely on existing inputs are inadequate. 
While automated test case generation techniques, such as symbolic
execution can be leveraged, adapting them to 
interrupt-driven software is not straightforward. 
For example, IIR is a read-only register and
thus cannot be directly manipulated; the value
of IIR is controlled by the interrupt enable register
(IER). Therefore, hardware  properties must be 
considered when generating input data. 


Fourth, races detected by static analysis or
symbolic execution without considering the change of
program control flow due to interrupts may report
false positives. In this example, static analysis would report there exists
a race on {\tt xmit->tail} between line 22 and line 31.
But the two lines can never get executed in the same run.
Therefore, controllability is needed
to validate whether a race detected by these techniques
are real or not. While several existing
approaches have tried to abstract away scheduling
non-determinism in concurrent programs to achieve
greater execution control (\eg,~\cite{Sen08}).
These approaches often control thread scheduling, but
cannot force hardware interrupts to occur at arbitrary point.

Finally, repairing interrupt-related race conditions requires
enforcing synchronization operations that are specific to interrupts.
In this example, the interrupt disable operation 
associated with
{\tt irq1\_handler} should be inserted before line 14. However, finding
the right place to insert interrupt operations is challenging
because the correctness of program semantics must be guaranteed.
In addition, the length of a critical section must be considered
because a long critical
section may lead to timing or performance violations.
Existing techniques on repairing thread-level concurrency faults
cannot be directly adapted to address this problem.

\subsection{Leveraging Virtual Platforms in Testing}
Virtual platforms such as Simics provide
observability and fine-grained controllability features
sufficient to allow test engineers to detect
faults that occur across the boundary
between software and hardware.
\Name{} takes advantage of many features readily available in many virtual
platforms to tackle the challenges of testing for race conditions
in interrupt-driven embedded software. Particularly, we can achieve the level of
observability and controllability needed to test such systems by utilizing the virtual
platform's abilities to interrupt execution without affecting the states of
the virtualized system, to monitor function calls, variable values and system states,
and to manipulate memory and buses directly to force events such as interrupts and traps.
As such, \Name{} is able to stop execution at a point of interest and force a
traditionally non-deterministic event to occur. Our system then monitors the effects
of the event on the system and determines whether there are any anomalies.

\subsection{Comparing to Thread-level Race Detection Techniques}
\label{subsec:thread}
Although interrupts are superficially similar to threads
(\eg, nondeterministic execution), the two abstractions have subtle semantic differences~\cite{regehr2007interrupt}. As such,
thread-level race detection techniques
\cite{Kahlon09,Marino2009,Bond2010,Effinger-Dean2012,pldi2012,icse2012-ballerina}
cannot be adapted to address interrupt-level race conditions.

First, threads can be suspended by the operating system (OS) and thus 
the insertion of delays (\eg, sleep or yield instructions) can be used to control
the execution of threads. The status of each thread is also visible 
at the application level.  However, interrupts cannot block -- 
they run to completion unless preempted by other higher-priority interrupts. 
The inability to block makes it impossible to use advanced OS services
for controlling the occurrences of interrupts in race detection. 
In addition, the internal states of interrupts
are invisible to tasks and other interrupt handlers because of the non-blocking characteristics. 
As such, it is impossible to use code instrumentation for checking the status of interrupts.

Second, 
threads typically employ symmetrical preemption relations -- they can preempt each other.
In contrast, tasks and interrupt handlers (\ie, task vs. ISR and ISR vs. ISR) 
have asymmetrical preemption relations.
Specifically, interrupts cannot be preempted by normal program routines; instead, they can
be preempted only by other interrupts with higher priority, and this can occur only when the current interrupt handler
is set to be preemptible. 
The asymmetric relationship between interrupt handlers and tasks
invalidates the happens-before relations 
served as the standard test for detecting thread-level races.
\wymm{
Because their computed happens-before relation will not be precise if an interrupt is triggered during their computation, but thread-level methods do not handle interrupts.
}

Third, the concurrency control mechanisms employed by interrupts
are different.
A thread synchronization
operation uses blocking to prevent a thread from passing 
a given program point until the synchronization resource becomes available. 
However, concurrency control in interrupts involves disabling
an interrupt from executing in the first place. This is done by either
disabling all interrupts or disabling specific interrupts that may
interfere with another interrupt or task. 
As such, 
thread-level techniques
that rely on binary/bytecode instrumentation~\cite{Yu12maple,Sen08} to
control memory access ordering between threads
cannot be used to control the occurrences
of hardware interrupts.
In contrast, interrupt-level race detection techniques
must be able to control hardware states
(\eg, registers) to invoke interrupts
at specific execution points~\cite{yu2012simtester}.
In addition, occurrences
of interrupts are highly dependent on hardware states;
that is, interrupts can occur only when hardware
components are in certain states. Existing thread-level race detection techniques
are not cognizant of hardware
states.

\begin{figure*}[!t]
\centering
\includegraphics[width=.95\textwidth]{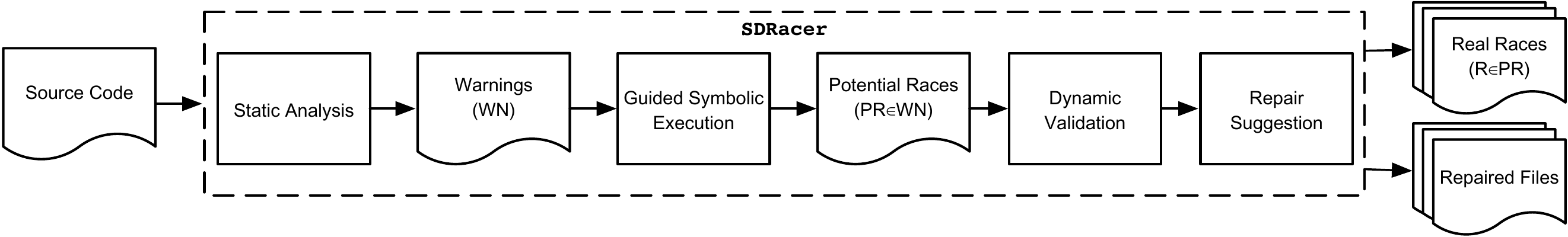}
\vspace*{-15pt}
\caption{\small Overview of \Name{} framework.}
\label{fig_sim}
\end{figure*}

\section{\Name{} Approach}
\label{sec:mod}

We introduce \Name{} whose architecture is shown in
Figure~\ref{fig_sim}.
The rectangular boxes contain the
major components.
\Name{} first employs lightweight static analysis (SA)
to identify potential sources of race conditions.
The output of this step is a list of static
race warnings, \{$<$$e_i$ = ($T_i$, $L_i$, $A_i$), 
$e_j$= ($T_j$, $L_j$, $A_j$)$>$\}. 
\wymm{However, the event pair $<e_i, e_j>$ is unordered and thus
\Name{} attempts to force
$e_j$ to occur after $e_i$ in the following components to validate the race order.} 
Here, $T$ is a task or an ISR, $L$ is the code
location, and $A$ is the access type.
In the example of Figure~\ref{exp1}, the output
of this step is:
$WN_1$ = $<$({\tt transmit}, 14, R), ({\tt irq1\_handler}, 22, W)$>$,
$WN_2$ =$<${\tt transmit}, 14, R), ({\tt irq2\_handler}, 31, W)$>$, 
$WN_3$ = $<${\tt irq2\_handler}, 31, W), ({\tt irq1\_handler}, 22, R)$>$,
and 
$WN_4$ = $<${\tt irq2\_handler}, 31, W), ({\tt irq1\_handler}, 24, R)$>$.

Next, \Name{} invokes symbolic execution to generate
input data that can reach
the code locations of the static race warnings.
In Figure~\ref{exp1}, 
the input data $t_1$ = \{{\tt IIR} = 0x0111, 
{\tt THR} = 0x0111, {\tt port->bugs = 0}\} is generated
to exercise $WN_1$, and $t_2$ = \{{\tt IIR} = 0x0111, 
{\tt THR} = 0x0110, {\tt port->bugs = 0}\} is generated
to exercise $WN_2$ and $WN_4$.
This step can also eliminate infeasible racing pairs. For example, 
$WN_3$  cannot be covered 
due to the conflict path conditions between
 {\tt irq1\_handler} and {\tt irq2\_handler}.
Therefore, $WN_3$ is a false positive.
The output of symbolic execution is a list
of potential races $PR$ and their corresponding input data. 


Then, \Name{} utilizes
the virtual platforms to exercise the inputs on the
potential races generated from the symbolic execution
and force the interrupts to occur at the
potential racing points.  The output of this step is a set of real races.
In the example of Figure~\ref{fig_sim}, $WN_1$
and $WN_4$ are real races because we can force the 
{\tt irq1\_handler} to occur right after line 14 and 
the {\tt irq2\_handler} to occur right after 24. 
Therefore, $WN_1$ and $WN_4$ are real
races, whereas $WN_2$ is a false positive; 
{\tt irq2\_handler} cannot be issued after line 14
because its interrupt line is disabled. 

\wymm{
Last, the race repair component repairs races
by enforcing the interrupt disable and enable operations, adding locks or extending critical sections.
It is worth noting that disabling interrupt will not lead the lost of interrupt because the interrupt is queued and will be saved to execute later\cite{corbet2005linux}.
In order to fix races, \Name{} first determines where to enforce these operations
and then generates patches by static code transformation.
In Figure~\ref{exp1}, to repair $WN_1$,
an interrupt disable operation 
{\tt irq\_disable(1)} is added right before line 14
to disable {\tt irq1\_handler}, and
an interrupt enable operation {\tt irq\_enable(1)}
is added right after line 14. 
Likewise, to repair $WN_4$, {\tt irq\_disable(1)} is added before line 31,
and {\tt irq\_enable(1)} is added right after line 31.
}

\subsection{Static Analysis}

In the static analysis phase,
\Name{} first identifies shared resources and interrupt
enable and disable operations. It then
analyzes a list of potential racing pairs,
\ie, static race warnings.
The racing pairs are used for
guiding symbolic execution and dynamic 
race validation.

\begin{algorithm}[t]
\small
\caption{Shared resources identification}
\label{alg:SRI}
\begin{algorithmic}[1]
\raggedright
\REQUIRE Task set $Tasks$, ISR set $ISRs$
\ENSURE shared resources set ($SRS$)

\STATE /* $S$ denotes all tasks and ISRs. $aliasSet$ is an alias set. */
\STATE $S$ = $Tasks \cup ISRs $, $aliasSet$ = $\emptyset$

\STATE /* Whole-program alias analysis */ \label{SRI:IA1} 
\FOR{ $T \in S$}
	\STATE $aliasSet$ = $aliasSet \cup$ andersenPointsToAnalysis($T$) 
\ENDFOR \label{SRI:IA2}

\STATE /* Construct inter-procedure aliases in $aliasSet$. */
\STATE $aliasSet$ = $aliasSet \cup$ linkAliasSet($S$, $aliasSet$) \label{SRI:LAS}

\FOR{each $ISR \in$ $ISRs$}
    \STATE /* $r$ denotes a variable in ISR */
	\FOR{each $r$ $\in$ $ISR$} 
	    \STATE /* $\mathcal{S}_{global}$ denotes all global variables, $parm(ISR)$ denotes parameters of $ISR$  */ \label{SRI:EXL1}
	    \IF{$r \notin  \mathcal{S}_{global}$ and $ \forall v \in parm(ISR), (v, r) \notin aliasSet$}
	        \STATE continue 
	    \ENDIF \label{SRI:EXL2}
	    \IF{$\exists T \in (S-\{ISR\}), r \in T$ or $alias \in T$ where $(r, alias) \in$ $aliasSet$, }
	        \STATE $SRS$ = $SRS \cup r$
	        \FOR{each $(r, alias)$ $\in$ $aliasSet$}
	                \STATE $SRS$ = $SRS \cup alias$ 
	        \ENDFOR
	    \ENDIF
  	\ENDFOR
\ENDFOR
\end{algorithmic}
\end{algorithm}

\subsubsection{Identifying Shared Resources}

Race conditions are generally caused by inappropriate synchronized
access to shared resources. So precisely detecting shared
resources is key to race detection.
In addition to shared memory that is considered by thread-level
race detection techniques,
\Name{} also accounts for hardware
components that are accessible by applications and device drivers,
including device ports and registers.

\Name{}~automatically decomposes tasks based on the specific patterns 
of device drivers. For example, the first parameter of $kthread\_create$
refers to the function name of a task. Another type of task is the function 
callback, which is  triggered by a specific device operation (\eg, device read).
\wymm{
Then, we use Algorithm \ref{alg:SRI} to identify all possible shared resources. 
Specifically, we use Andersen's pointer analysis \cite{andersen1994program} to 
identify resources accessed by at least: 
1) two ISRs, or 2) one task and one ISR.
For each detected shared resource, we also add all its aliases to the shared resource set $SRS$.
We first perform intra-procedure alias analysis for each task and ISR (at lines \ref{SRI:IA1} to \ref{SRI:IA2}).
Then, in order to connect alias relation in tasks/ISRs, we assume interrupt request operations (\eg, $request\_irq$) and task registration operations (\eg, $kthread\_create$) is called right after its registration.
For example, $threadfn(threadArg);$ and $interruptFN(dev)$ are called right after $kthread\_create(threadFn, threadArg, ...);$ and $request\_irq(irqNum, interruptFn, flag, ...);$ separately.
Based on the assumption, we perform inter-procedure alias analysis (at line \ref{SRI:LAS}).
In this step, we analyze the program from top to bottom according to its call graph.
For each function, when a variable is used as an argument of a callee, we combine its aliases and aliases of the argument in the callee.
Therefore, a caller's alias relation will be passed to its callees iteratively, and thus the assumption will not miss possible aliases.
Then, we identify shared resources based on the whole program alias relation.
For each ISR, we only analyze a variable (\ie, $r$) in the ISR if it is a global variable (including hardware components like device ports) or it may be an alias of the ISR's parameter (at lines \ref{SRI:EXL1} to \ref{SRI:EXL2}).
Finally, for each remaining variable, we add it and its aliases to $SRS$ if 1), it is also used in other tasks or ISRs (\ie, $r \in T$ denotes $r$ is a global variable), or 2) at least one of its aliases may be used in other tasks or ISRs (\ie, $alias \in T$ where $(r, alias) \in$ $aliasSet$).
}

Each detected shared resource access $SRA$ is denoted by a
6-tuple: $SRA$ = \{$<T, L, V, AV, R, A>$\}, 
where \emph{T} denotes
the name of the task or ISR in which the shared resource ($SR$) is accessed,
\emph{L} denotes the code location of the access,
\emph{V} denotes the name of the $SR$,
\emph{AV} denotes whether the name $V$ is an alias ($false$) or
a real name ($true$) (real name is the declared name),
\emph{R} means the real name of this
resource, and \emph{A} denotes the access type -- read (denoted by R)
or write (denoted by W).

In the example of Figure~\ref{exp1}, all $SRA$ for
the {\tt xmit$\rightarrow$tail} is: \\
$<${\tt transmit}, 14, {\tt xmit$\rightarrow$tail}, 
true, {\tt xmit$\rightarrow$tail}, R$>$,
$<${\tt transmit}, 22, {\tt xmit$\rightarrow$tail}, 
true, {\tt xmit$\rightarrow$tail}, W$>$,
$<${\tt irq\_handler}, 24, {\tt xmit$\rightarrow$tail}, 
true, {\tt xmit$\rightarrow$tail}, R$>$, and
$<${\tt transmit}, 31, {\tt xmit$\rightarrow$tail}, 
true, {\tt xmit$\rightarrow$tail}, W$>$.

\subsubsection{Identifying Interrupt Operations}
To track interrupt status (\ie, disabled or 
enabled) of a shared resource, \Name{}
identifies interrupt-related synchronization operations,
which typically involve interrupt 
disable and enable operations.
In many embedded systems, coding interrupt
operations can be rather flexible.
An interrupt operation can be done by directly manipulating
hardware bits (\eg, line 9 of Figure~\ref{exp1}).
In addition, these operations vary across
different architectures and OS kernels.

\Name{} considers both explicit and implicit
interrupt operations. 
For the explicit operations,
\Name{} considers standard Linux interrupt APIs,
including {\tt disable\_irq\_all()}, {\tt disable\_irq(int irq)}, {\tt
disable\_irq\_nosync(int irq)} and {\tt enable\_irq(int irq)}, where the {\tt irq} parameter
indicates the interrupt vector number (\ie, the unique
ID of an interrupt).  
For the implicit operations, \Name{} tracks operations
that manipulate interrupt-related hardware components,
such as the interrupt enable registers ({\tt IER}s). 
\wymm{Since the effect of these operations are often not recognized by static analysis,}
\Name{} conservatively assumes they are equivalent to
interrupt enabling (\eg, {\tt enable\_irq\_all()}) to avoid false negatives.
The consequence is false positives, which can be validated by the dynamic validation. 
In Figure~\ref{exp1}, the hardware write operation
at line 9 is considered to be an interrupt enable operation.

To handle interrupts in different kernels
or architectures. \Name{} provides
a configuration file that allows developers to
specify the names of interrupt APIs.
The output of this step
is a 4-tuple list: ITRL = \{$<$$M, L, I, T$$>$\}, where \emph{M}
denotes the function name, \emph{L} denotes the code location
where the interrupt operation is called,
\emph{I} denotes
the interrupt vector number and \emph{T} denotes the type of
interrupt operation (\ie, enable or disable).
In the example of Figure~\ref{exp1}, the $ITRL$ is:
$<${\tt transmit}, 9, $all$, $enable$$>$,
where $all$ denotes all interrupts are enabled.

\begin{algorithm}[t]
\small
\caption{Static race detection}
\label{alg:idenSRs}
\begin{algorithmic}[1]
\raggedright
\REQUIRE RICFGs of $P$
\ENSURE potential racing pairs ($PR$)
\FOR{each $<G_i, G_j>$ in $RICFGs$}
	\FOR{each $sv_i$ $\in$ $G_i$}
		\FOR{each $sv_j$ $\in$ $G_j$}
			\IF{$sv_i.V$ == $sv_j.V$ and ($sv_i.A$ == $W$ or
			$sv_j.A$ == $W$) and $G_i.pri$ $<$ $G_j.pri$ and  INTB.{\tt get}($sv_i$).{\tt
			contains}($G_j$)} 
			\STATE $PR$ = $PR \cup (sv_i, sv_j)$
  			\ENDIF
  		\ENDFOR
  	\ENDFOR
\ENDFOR
\end{algorithmic}
\end{algorithm}

\subsubsection{Identifying Static Race Warnings}

In this step, we identify shared resource access pairs
that may race with each other from all identified shared resources.
These pairs are used as targets for guiding symbolic execution
to generate test input data.

%
To statically identify potential racing pairs,
we first build a reduced inter-procedural control flow
graph (RICFG) for the task and each of the ISR
that contains at least one shared resource. 
RICFG  prunes branches that do 
not contain shared resources in the original
inter-procedural control flow
graph (ICFG)
in order to reduce the cost of analysis. 
\wymm{Additionally, we use a bit vector \emph{INTB} to record the interrupt
status. For example, \emph{INTB} = $<$1, 0, 0$>$ indicates 
that the first interrupt is disabled and the second and
the third interrupts are enabled. 
\emph{INTB} is updated when an interrupt disable/enable instruction is visited.  Note that when 
visiting an instruction inside the ISR, the bit associated with
the ISR is always set to 1 because an ISR is non-reentrant.
Additionally, it is possible that an interrupt disabled by a task or an ISR is re-enabled immediately by another interrupt.
In order to avoid false negatives, for each interrupt disable instruction, we check all enabled interrupts right after this operation. 
If there exists an interrupt that re-enables the disabled interrupt, we ignore the interrupt disable operation and its corresponding interrupt enable operation.
By this conservative analysis method, we can avoid disabling interrupts that can be enabled by other ISRs.

Algorithm~\ref{alg:idenSRs} describes the computation
of potential racing pairs based on the 
RICFGs of the program. 
\Name{} traverses each RICFG by a
depth-first search to examine 
the interrupt status (\ie, enable or disable)  of every instruction. 
} 
For each shared resource $sv_i$ at the location
$L$ of a RICFG $G_i$, if there exists the same
shared resource $sv_j$ in a RICFG $G_j$,
at least one shared resource is a write, the priority
of $G_j$ is higher than that of $G_i$, and
the interrupt for $G_j$ is enabled at $L$,
the pair ($sv_i$, $sv_j$) forms a potential
race condition. For example, in Figure~\ref{exp1},
the bit vector at line 14 is $<$0, 0$>$, indicating
that both irq1 and irq2 are enabled.  Also, both irq1 and irq2 
have higher priorities than {\tt transmit}. 
The bit vector at line 13 is $<$1, 0$>$,
because {\tt irq1\_handler} is non-reentrant. 
Therefore, $WN_1$, $WN_2$, $WN_3$, and $WN_4$ are 
reported as static race warnings.

\wym{About loops, our lightweight static analysis does not fully analyze
loops because we unroll loops twice (\ie, transform a loop into two if statements) to balance accuracy and efficiency.
Moreover, the analysis is context-insensitive, which may lead to false positives
because it does not distinguish between different calling contexts of a 
function call. 
On the other hand, precise static analysis is 
more expensive~\cite{Whaley2004}.
As future work, we will evaluate cost-effectiveness 
by adopting precise static analysis techniques.
}

\subsection{Guided Symbolic Execution}

We propose a new symbolic execution procedure to generate
input data for exercising static race 
warnings reported in static analysis and eliminating
a portion of false races. 
Unlike traditional guided symbolic execution~\cite{Marinescu12,Farzan13},
symbolic execution on interrupt-driven programs
needs to consider the asymmetrical preemption relations
among tasks and ISRs. 
The symbolic execution of \Name{} consists of two steps:
1) identifying entry points that take symbolic inputs;
2) generate inputs that exercise
racing pairs reported by static analysis.
Internally, we leverage the KLEE symbolic virtual machine~\cite{Cadar08} to
implement the goal-directed exploration of the program to traverse the program
locations involving potential races.

\subsubsection{Identifying Input Points}
Execution paths in embedded systems usually depend on 
various entry points that accept inputs
from external components, such as registers 
and data buffers~\cite{yu2016vdtest}. 
One challenge for our approach involves
dealing with multiple input points in order to 
achieve high coverage of the targets. 
\Name{} considers two kinds of input points:
1) hardware-related memories (\eg, registers, DMA), and
2) global data structures used to pass across
components (\eg, buffers for network packages,
global kernel variables that are accessible
by other modules). 
\Name{} can identify these input points based 
on the specific patterns of device drivers -- this
is a per-system manual process.

In the example of Figure~\ref{exp1}, The input points
include the UART registers and the UART port. 
Specifically, the values in the registers IIR (line 3)
and THR (line 21 and line 30) determine
the data and control flow of the program execution.  
As such, we make these register variables symbolic. 
We also make the data fields of the UART port 
symbolic (\eg, {\tt port->bug} at line 4)
because they accept inputs from users and external 
components.

\subsubsection{Guided Symbolic Execution.}
For each static race warning $WN$ = $<$$e_i$, $e_j$$>$,
\Name{} calls the guided symbolic execution
to generate a test input to exercise the $WN$
or report that the $WN$ is a false positive.
Since each call to the symbolic execution targets
a pair of events in two different tasks or ISRs,
we build an inter-context 
control flow graph (ICCFG) by connecting the 
inter-procedural control flow graphs (ICFGs)
of the tasks and ISRs. 
For an instruction that is equal to the
first racing event $e_i$ in a $WN$,
we add an edge that connects $e_i$ to the entry function of
the ICFG in which $e_j$ exists. 
In the example of Figure~\ref{exp1}, to generate inputs for 
$WN_1$ = $<$({\tt transmit}, 14, R), ({\tt irq1\_handler}, 22, W)$>$,
the entry of {\tt irq1\_handler} is connected to 
the instruction right after the {\tt xmit->tail} read access. 

\wym{
\Name{} guides the symbolic execution toward
the two ordered events of each WN by exploring 
the ICCFG. 
Let $e \in WN$ denote the current event to
be explored, and $s$ denotes the current program state that symbolic execution is exploring. 
A program state \cite{cadar2008klee} is a representation of symbolic process.
For each branch and loop, the program state will be cloned to explore different paths.
Based on $s$, we refer to $\mathcal{S}_s$ as the set of next program states that symbolic execution could explore and reach $e$.
Given $e$ and $s$, $\mathcal{S}_s$ can be analyzed by the backward reachability analysis of ICCFG.  
At each step of the symbolic execution procedure, we select a promising state
$s_i \in \mathcal{S}_s$, which is likely to reach $e$. 
Internally, \Name{} estimates the distance between each program state
$s_i$ and $e$ before selecting the next state. 
Note that $s_i$ is a next program state cloned based on branches or loops.
Therefore, each $s_i$ contains the next instruction to be explored.
\wymm{We define the distance from $s_i$ to $e$ to be the minimum number of instructions from the next instruction in $s_i$ to the instruction in $e$ according to the ICCFG.}
Among all program states in $ \mathcal{S}_s$, we select $s_i$ to be the next program state that symbolic execution should explore if its next instruction leads to the shortest path to the target instruction.
}
If multiple states have the same distance to $e$,
\Name{} randomly selects one.
In this sense, the search strategy of \Name{} differs 
from prior symbolic execution techniques such as state prioritization
(\eg, assertion-guided symbolic execution~\cite{Guo15} and
coverage-guided symbolic execution~\cite{Kuznetsov2012,Cadar08, ma2011directed}),
because they do not target the exploration of
potential racing points.

If no state in $stateset$ can reach $e$, we check if $e$ is in a
loop.  If $e$ is in a loop, we increase the number of loop
iterations by a fixed number of times given a timeout threshold and
try again. This will increase our chance of reaching the goal.
The iteration number is increased until reaching the loop bound $L_{max}$
($L_{max}$ = 1000 in our experiments).

Otherwise, we backtrack and search for another
path to the current event.  If backtracking is repeated many times,
eventually, it may move back to the first event, indicating that the
current racing pair cannot be exercised. In such case, 
we move to the next racing pair.
%
\wym{
When our symbolic execution reaches the first event (\ie, $e_i$), the next program state to explore is the entry instruction in $e_j$.
We continue exploring program states until we reach the second event.
After reaching the second event (\ie, $e_j$), we traverse the current program
path to compute the path condition (PC).  Then, we compute the
data input by solving the path condition using an SMT
solver.  
}

The main problem in guided symbolic execution is to make the procedure
practical efficient by exploring the more ``interesting'' program
paths.  Toward this end, we propose several optimization techniques.
\wym{
Recall the way of constructing RICFG that we statically analyze the source code of the program to prune away paths that do not reach the shared resources -- they are
irrelevant to the potential races.  
}
We also skip computationally expensive constraint solver calls unless
the program path traverses some unexplored potential races.
In addition to these optimizations, we prioritize the path exploration
based on the number of potential races contained in each path
to increase the likelihood of reaching all static races sooner.
%
\wym{Furthermore, we leverage concrete inputs (randomly generated) to help solving complex path constraints.}


%
In the example of Figure \ref{exp1}, the symbolic
execution successfully generates
input data for exercising $WN_1$ 
and $WN_2$, and  $WN_4$.
For $WN_3$, the symbolic
execution explores the two events at  
line 31 and line 21
in the ICCFG that connects {\tt irq1\_handler} and 
{\tt irq2\_handler}. 
The path constraint  
$ thr == 0x1101 \wedge thr \neq 0x1101$
is unsolvable, so $WN_3$ is a false positive. 
%

For each static warning, 
there are three types of output generated by
the symbolic execution. 
The first type of output is a potential
race together with its input data, 
which means that this race is possible to 
be exercised at runtime. 
The second type of output is an
unreachable message
(unsolvable path constraints), which indicates
that the static warning is a false positive. 
The third type of output is a message
related to timeout or crash. 
The reason could be the execution
time-out, the limitation of constraint solver 
or the unknown external functions. 
In the next phase of dynamic validation,
we validate whether races reported 
in the first and third types are real
races or not.

\subsection{Dynamic Validation of Race Conditions}
\label{sec:DYN}

We propose a hardware-aware dynamic analysis method to validate
the remaining race conditions from the symbolic execution.
In this phase, \Name{} simulates virtual environment for interrupt-driven programs, which provides an execution observer and an execution controller.
First, it employs an execution
observer to monitor shared resource accesses
and interrupt operations, and then uses
an execution controller to force
each race condition to occur.

\subsubsection{Executing Observer}
The Observer records operations
that access shared memory and hardware
components. The observer also 
monitors interrupt bits (IER and IIR registers) to track 
interrupt disabling and enabling operations. 
These bit-level operations are then mapped 
into the instruction-level statement, because 
the control of interrupts happens at the instruction level. 

For each shared resource access, \Name{}
can retrieve the current interrupt status
of all IRQ lines to check whether it is possible to force 
a specific interrupt to occur.

\begin{algorithm}[t]
\small
\caption{Execution controller}
\label{fig:alg3}
\begin{algorithmic}[1]
\raggedright
\REQUIRE $PRaceSet$, $P$, $S$
\ENSURE $RaceSet$
\FOR{each $\sigma$ = ($e_{i}$, $e_{j}$) $\in$ $PRaceSet$}
	 \IF{$e_{i}$ in $T$}
	 	\STATE $E$ = {\tt Execute($P$, $t_\sigma$)} 
	 \ENDIF  
	 \IF{$e_{i}$ in $H$}
	 	\STATE $E$ = {\tt Execute($P$, $e_{i}.H$, $t_\sigma$)} 
	 \ENDIF 
	 \IF{$E$ covers $e_{i}$}
        \IF{{\tt ISR\_enabled ($e_{j}$.H)} is true}        
  				\STATE raise interrupt $e_{j}.H$
  		\ELSE
  			    \STATE find another possible location 
  	    \ENDIF
  	    \IF {$e_{j}.H$ accesses $e_{j}$}
  	    	\STATE $RaceSet$ = $RaceSet \cup \sigma$ /*race occurs*/ 
  	    \ENDIF
  	    \IF {{\tt Output($P$, $S$)} $\neq$ $O$}
  	    	\STATE print ``Error: fault found'' 
  	    \ENDIF
  	 \ENDIF
\ENDFOR
\end{algorithmic}
\end{algorithm}

\subsubsection{Execution Controller}
\label{subsub:isr}

Simics allows us to issue an interrupt 
on a specific IRQ line from the simulator itself. 
As such, when the Observer reaches an $SV$, an interrupt
is invoked at a feasible location after the access
to this $SV$.  

We now describe the algorithm of
execution controller (Algorithm \ref{fig:alg3}).  
Given a potential racing pair $\sigma$ = ($e_{i}$, $e_{j}$),
The goal of this algorithm is to
force an ISR that contains $e_{j}$ to occur 
right after the access to $e_{i}$.
The algorithm first executes the program under test $P$ (line~6).  
If the the first shared resource access $e_i$ occurs
in a task, the algorithm executes the input data
(generated from the symbolic execution) on $P$ (line 3).
If $e_i$ occurs in an ISR, it executes $P$ together
with the interrupt issued at the arbitrary location
of $P$ (line 6). If the execution covers $e_i$,
the algorithm  forces the interrupt 
in which $e_j$ exists to occur immediately after 
$e_i$ (line 9).  
If a race occurs, it is added to $RaceSet$ (line~15). 

Note that our algorithm can also force the interrupt 
to trigger immediately before $e_i$. 
In fact, the effect of triggering an interrupt immediately after the first event
covers that of triggering an interrupt before the first event because 
a failure is usually caused by reading the incorrect value modified by 
the interrupt handler. 
\wymm{
It is not critical to choose either case because if we can trigger the second event right after the first event, then we can also trigger the second event right before the first event.
}

Because it may not be possible to raise an interrupt
immediately (\eg, if the interrupt is currently disabled), the
algorithm checks the current state of the interrupt
associated with $e_j$ (line 9) before raising an interrupt. 
The algorithm also checks outputs on termination of the
events (lines~17-18) to determine whether a fault
has been identified.
If the interrupt ($S$) cannot be raised
\emph{immediately after} the shared resource access
$e_i$ in $P$
(lines~9-10), the algorithm postpones $e_i.H$ (the ISR in which $e_j$
exists) until it can
feasibly be raised, or until the entry instruction of the
operation in another potential race pair is reached.

To illustrate the algorithm's operation, using
Figure~\ref{exp1} as an example.
Considering $WN_1$, given the input $t_1$,
the {\tt transmit}  covers the
read of {\tt xmit$\rightarrow$tail} at line 14. 
Thus, the algorithm forces {\tt irq1\_handler} to be 
raised right after the read of {\tt xmit$\rightarrow$tail} at line 14,
In this scenario, {\tt xmit$\rightarrow$tail} is modified
by the  {\tt irq1\_handler}, causing {\tt transmit} to 
read the wrong value.
As a result, $WN_1$ is real and harmful.

It is not 
always realistic to invoke an interrupt whenever we want. 
For example, the interrupt enables register 
and possibly other control registers have to
be set to enable interrupts. 
In the example of Figure~\ref{exp1}, 
before invoking an interrupt, 
the interrupt enable
register IER of the UART must be set 
while the interrupt identification 
register IIR must be cleared.  
Interrupts can be temporarily disabled even if they are enabled.
Algorithm~\ref{fig:control} is the routine in the Controller
used to determine whether it is possible to issue an interrupt. 

\begin{algorithm}[t]
\small
\caption{Algorithm to determine
whether it is possible to issue an interrupt: \textit{ISR\_enabled(int p)}}
\label{fig:control}
\begin{algorithmic}[1]
\raggedright
\REQUIRE P, p /* p is the pin number for a certain interrupt */
\ENSURE enabled
\IF{$eflags$[$9$] != 0 and {\tt ioapic.redirection[$p$]} == 0
   and {\tt ioapic.pin\_raised[$p$]} == $LOW$}
	\RETURN true
\ENDIF
\RETURN false
\end{algorithmic}
\end{algorithm}

There are two general steps that our system takes prior
to invoking a \emph{controlled interrupt}.  First, the
controller module checks the status of the local
and global interrupt bits to see if
interrupts are enabled. In an X86 architecture, the
global interrupt bit is the ninth bit of the eflags
register (line 1 in Algorithm~\ref{fig:control}).  
When this bit is set to~1 the global interrupt is disabled,
otherwise it is enabled.  For local interrupts, Simics
uses the Advanced Programmable Interrupt Controller
(APIC) as its interrupt controller. As such, our system
checks whether the bit controlling the UART device is
masked or not. 

\subsection{Race Conditions Repair Suggestions}

Once a race condition is validated, the next step is to 
repair it. 
Before designing repair suggestions for race condition,
we conduct an empirical study about the real-world repair methods for data race and race condition.

\subsubsection{Learning from practice}
The repair of race condition requires strong domain knowledge and technical background. 
At present, it relies mainly on the manual repair by developers, which is time-consuming and laborious, and is also prone to errors. 
In order to explore and summarize the successful repair patterns that can be reused or can be automated, we conduct a lightweight empirical study to analyze how developers repair the race condition.
Then we summarize the repair insight and patterns, in order to guide the developers for their future race repair as a reference.

To survey the repair strategies, we choose two data sources, the industry collaborators for embedding programs and the Linux kernel community.
The industry collaborators provide their commonly used repair strategies in interrupt-driven programs.
About collecting data in the Linux kernel, we inspect 532 bugs collected by Shi et al. \cite{Shi2018Linux} from 2011 to 2015 and validate 387 of them. 
\wym{Among the 387 races, 59 of them are related to interrupts since a part of races in the kernel are task-level races.}
The remaining bugs are not race conditions or too complex to identify their repair strategies.
The reason for choosing Linux kernel as the subject of the empirical study is that it is the closest project that consists of many interrupt-driven programs (\eg, drivers).

Table \ref{tab:sum} shows the result of race condition repair strategies.
For each repair strategy, we describe how the method achieves eliminating races (column 2), along with examples (column 3).
The fourth column indicates whether this type of repair strategy is used by our industry collaborators.
The fifth and sixth columns are collected from the Linux kernel, 
we provide the number of occurrence for each repair method and the corresponding percentage among task-level races (\ie, \textit{Task}) and interrupt-level races (\ie, \textit{Int}).
The last column discusses the conditions of applying repair suggestions.
\textit{NA} in the last row denotes that the two fields are not applicable. 

\begin{table*}[!t]
\centering
\small
\caption{Summary of repair strategies from the industry and Linux kernel}
\begin{tabular}{|c|c|c|c|c|c|c|}
\hline
\multirow{2}{*}{Repair strategy}   & \multirow{2}{*}{Description} & \multirow{2}{*}{Example} & \multirow{2}{*}{\tabincell{c}{From\\ind.}}  &  \multicolumn{2}{c|}{\tabincell{c}{From Linux}} &  \multirow{2}{*}{\tabincell{c}{Application\\condition}}  \\
\cline{5-6}
&  &   &  & Task & Int & \\\hline
\tabincell{c}{Change operation\\orders\\(COO)} & \tabincell{c}{Change the  order of\\operations so that the\\racing operations happen\\in separate timing\\$e_i \wedge e_j = false$}
  & \tabincell{c}{Move codes to\\a position where\\interrupts are \\finished} & \cmark & \tabincell{c}{88\\26.8\%} & \tabincell{c}{17\\28.8\%} &  \tabincell{c}{The separate timing\\is available} \\\hline 
\tabincell{c}{Add additional\\checks\\(AAC)} & \tabincell{c}{Add additional checks to\\check program states \\to avoid race\\$e_i \wedge e_j = false$}   & \tabincell{c}{if (!dev\_initialized())\\ wait\_until\_init();} & \xmark & \tabincell{c}{85\\25.9\%} &  \tabincell{c}{5\\8.5\%} &  \tabincell{c}{There is an available\\and race-free\\program state}   \\\hline
\tabincell{c}{Add locks\\(AL)} & \tabincell{c}{Add additional locks\\and unlocks\\$e_i \wedge e_j = false$} & \tabincell{c}{spin\_lock/\\spin\_unlock} & \checkmark & \tabincell{c}{81\\24.7\%} &  \tabincell{c}{0\\0\%} &  \tabincell{c}{It will not\\introduce deadlocks} \\\hline
\tabincell{c}{Interrupt disable\\and enable\\(IDE)} & \tabincell{c}{Disable and\\enable interrupts\\$disable \wedge e_j = false$} & \tabincell{c}{disable\_irq/\\enable\_irq} & \cmark &  \tabincell{c}{0\\0\%} & \tabincell{c}{26\\44.1\%} &    \tabincell{c}{It will not\\introduce deadlocks}      \\\hline
\tabincell{c}{Add atomic\\instructions\\(AAI)} & \tabincell{c}{Add atomic instructions\\$e_i \wedge e_j = false$}    & atomic\_set & \xmark & \tabincell{c}{19\\5.8\%} & \tabincell{c}{4\\6.8\%} & \tabincell{c}{Its corresponding\\atomic instruction\\is available} \\\hline
\tabincell{c}{Synchronization\\(Sync)} & \tabincell{c}{Synchronization\\$e_i \wedge e_j = false$} & \tabincell{c}{Read-copy update,\\memory barrier} & \cmark & \tabincell{c}{23\\7.0\%} &  \tabincell{c}{0\\0\%} &    \tabincell{c}{It should be used \\judiciously to avoid \\impeding performance}     \\\hline
\tabincell{c}{Remove\\race codes\\(RRC)} & \tabincell{c}{Remove race codes\\$remove \; e_i \; or \; e_j$} 	     & \tabincell{c}{Remove unnecessary\\but buggy codes} & \xmark & \tabincell{c}{12\\3.7\%} & \tabincell{c}{2\\3.4\%} & \tabincell{c}{The racy code is\\no longer needed}\\\hline 
\tabincell{c}{Extend critical\\sections\\(ECS)} & \tabincell{c}{Extend critical sections\\$e_i \wedge e_j = false$}    & \tabincell{c}{Move spin\_unlock\\after the racing code.} & \xmark & \tabincell{c}{10\\3.0\%} & \tabincell{c}{4\\6.8\%} &  \tabincell{c}{It will not\\introduce deadlocks}    \\\hline
\tabincell{c}{Minimize the use\\of shared resources\\(MinUse)} & \tabincell{c}{Minimize the use\\of shared resources\\$m_i \neq m_j$} & \tabincell{c}{Use bit operation\\instead of value\\assignment} & \xmark & \tabincell{c}{3\\0.9\%} &  \tabincell{c}{0\\0\%} &    \tabincell{c}{Some SV accesses\\are redundant}     \\\hline 
\tabincell{c}{Add try-again\\marks\\(ATM)} &  \tabincell{c}{Retry interrupted tasks \\$\exists e_i \wedge e_j = false$}     & \tabincell{c}{T()\{if(flag==0)...\}\\ISR()\{flag=1;...\}} & \cmark &  \tabincell{c}{2\\0.6\%} &  \tabincell{c}{0\\0\%} &   \tabincell{c}{Performance\\insensitive tasks\\or interrupt handlers}      \\\hline
\tabincell{c}{Restrict users\\(ResUser)} & \tabincell{c}{Restrict users by\\documents or user manual\\$e_i \wedge e_j = false$} & \tabincell{c}{Forbid user sending\\requests right after\\starting a device.} & \cmark  & \tabincell{c}{0\\0\%} &  \tabincell{c}{0\\0\%} &   \tabincell{c}{A general method}  \\\hline
\tabincell{c}{Change priorities\\of tasks or\\interrupts\\(ChgPrio)} & \tabincell{c}{Change priorities of tasks\\or interrupts\\$p_j > p_i = false$} & \tabincell{c}{Reverse priorities\\of two interrupts} & \cmark  & \tabincell{c}{0\\0\%} &  \tabincell{c}{0\\0\%} &  \tabincell{c}{It will not\\lead to other races}  \\\hline
\tabincell{c}{Ad hoc repairs\\(Others)} & \tabincell{c}{Mostly $e_i \wedge e_j = false$} & NA & \xmark &  \tabincell{c}{5\\1.5\%} &  \tabincell{c}{1\\1.7\%} & NA  \\\hline
\end{tabular}
\label{tab:sum}
\end{table*}%

\subsubsection{Repair suggestions}

As shown in Table  \ref{tab:sum}, in order to repair the race condition, developers can add code snippets, remove code snippets, and modify code snippets at specific locations to eliminate the root causes of the race condition.

In real-world repair, the race condition repair requires the location of shared resources, the location of their read and write operations, the interleaving of interrupts and tasks, which are very complicated. 
Moreover, in some embedded systems, the race condition repair method may consider the state of devices, in order to give a specific and ad hoc fix.

As far as we know, it is impossible to fully automate the repair process without programmers' participation \cite{khoshnood2015concbugassist}.
Therefore, developers can choose repair strategies according to Table \ref{tab:sum} and repair the program manually by themselves or automatically by \Name{}. 

Based on Table \ref{tab:sum}, we can see that only a few of them are general purpose repair strategies, including \texttt{AL}, \texttt{AAI}, \texttt{IDE}, \texttt{Sync} and \texttt{AMB}. Others depend on semantics and pattern of the race.
Among the top 5 repair strategies,
\texttt{AL} and \texttt{IDE} are applicable to automatic repair.
In comparison, {\tt COO} and {\tt AAC} depend on the semantics of the program and thus are more difficult to be done fully automatically.
{\tt AAI} only provides a limited set of operations, and often these operations are not enough to synthesize more complicated operations efficiently.
\wym{It is worth noting that among all interrupt-level races, 44.1\% of them are fixed by \texttt{IDE}.}
Among the remaining strategies, \texttt{Sync} and \texttt{AMB} are relatively uncommon and others depend on the pattern of races.
Therefore, we focus on \texttt{AL} and \texttt{IDE} considering difficulty, practicality and universality.
We leave other repair strategies as future work.


\textbf{Interrupt disable and enable strategy (\texttt{IDE}):}
This strategy automatically enforces
interrupt disable and enable operations (\eg, 
\texttt{disable\_irq(int irq)} and 
\texttt{enable\_irq(int irq)}) on tasks 
or ISRs to avoid triggering the interrupt
that can result in races. 
Note that we only disable and enable the interrupt line leads to races.
Moreover, without loss of generality, we assume the interrupt disable operation waits for any pending IRQ handlers for this interrupt to complete before returning \cite{disirqdoc}. 

\wymm{
The main challenge of applying disabling/enabling interrupts strategy is that improper use of interrupt disable operation may lead to deadlock. 
For example, when a task disables an interrupt while holding a mutual exclusion shared resource (\eg, spin\_lock\cite{spinlockdoc}) the interrupt handler also needs to hold. Then the interrupt has to wait for the resource, but it leads to deadlock since the task that holds the resource also has to wait for the interrupt to return.

Given a racing pair (\{$<$$e_i$ = ($T_i$, $L_i$, $A_i$), 
$e_j$= ($T_j$, $L_j$, $A_j$)$>$\}. ), 
Equation~\ref{equ:deri1} is a sufficient condition to insert an interrupt disable operation right before instruction $I_d$ while avoiding this kind of deadlock. 
Because there is no mutual exclusion shared resource between $I_d$ and $T_j$.
$I_d$ is an instruction in $T_i$, indicating the location of inserting an interrupt disable operation. 
$\mathcal{I}_{T_j}$ is a set of all instructions in $T_j$.
$hold(i)$ denotes all possible mutual exclusion shared resources (\eg, spin\_lock) that instruction $i$ may hold and not released. 
$hold(i)$ is calculated based on the alias analysis in Algorithm \ref{alg:SRI}.
First, we analyze all mutual exclusion shared resources that instruction $i$ holds, and add it to $hold(i)$.
For example, a spin\_lock is in $hold(i)$ if and only if it is acquired before $i$ and released after $i$ according to RICFG.
Second, for each resource in $hold(i)$, we analyze its aliases based on the alias analysis result and add them to $hold(i)$.



\begin{equation} \label{equ:deri1}
hold(I_d) \cap (\bigcup_{k \in \mathcal{I}_{T_j}} hold(k) ) = \emptyset 
\end{equation}

Then, let $Preds(i)$ be all predecessors of the instruction $i$ (inclusive) in RICFG and let $I_i$ be the instruction at $L_i$.
Based on the equation, we identify $I_d$ if $I_d \in Preds(I_i)$, $I_d$ meets Equation~\ref{equ:deri1} and $\forall I_j \in Preds(I_i), dis(I_d, I_i)  \leq dis(I_j, I_i)$. $dis(i, j)$ is the minimum number of instructions from instruction $i$ to $j$ in RICFG.

After locating $I_d$, to identify the instruction of inserting interrupt enable operation (denoted as $I_e$), we analyze $T_i$ to find an instruction $I_e$ \st $\mathcal{S}_d = \text{post-dom}(I_d), \mathcal{S}_e = \text{dom}(I_e)$ in RICFG, $I_d \in \mathcal{S}_e$, $I_e \in \mathcal{S}_d$, and $\forall I_k \in \mathcal{S}_d, dis(I_e, I_d)  \leq dis(I_k, I_d)$. In other words, $I_e$ post-dominates $I_d$, $I_d$ dominates $I_e$ and $I_e$ is the closest location to $I_d$. If we cannot find $I_e$ or $I_d$, then we report this race pair is unable to repair by this strategy.
Finally, we insert an interrupt disable instruction right before $I_d$ and an interrupt enable instruction right after $I_e$. 
}

In the example of Figure \ref{exp1}, 
to repair the race $WN_1$,
\Name{} first inserts an \texttt{disable\_irq(1)}
right before line 14. 
Then the  \texttt{enable\_irq(1)} is inserted
right after line 14.  

\textbf{Add locks strategy (\texttt{AL}) and extend critical sections (\texttt{ECS}):}
The core idea of this repair strategy is to automatically create or extend a critical section that protects shared variables. 
The principle is that keeping the order of existing locks will not introduce deadlock problem \cite{cai2017adaptively}.

\wymm{
First, we define the lock order. 
If a task or an ISR firstly acquires a lock $l_i$ and then acquires another lock $l_j$, there is a lock order from lock $l_i$ to lock $l_j$, denoted by $(l_i, l_j)$.
There is a special case that, if a task or an ISR acquires two locks $l_m$ and $l_n$ at the same time (\eg, aliases), then the two locks have the same lock order.
For example, if $l_m$ and $l_n$ are acquired at the same time after acquiring $l_j$, then the lock order is $(l_i, l_j, l_m/l_n)$.
The lock order is computed by traversing the IRCFG of a program, recording each lock acquire operation.
Given a racing pair ($e_i$, $e_j$), assume the order of acquiring locks for $e_i$ is $LS_i = (l_{i1}, ... , l_{ik})$. 
Similarly, the relation for $e_j$ is $LS_j = (l_{j1}, ... , l_{jm})$. 
Rule~\ref{equ:deri2} is a sufficient condition to identify deadlocks because it avoids any inconsistent lock order between $LS_i$ and $LS_j$. 
$l_i \rightarrow l_j$ denotes that lock $l_i$ is acquired before holding $l_j$.
Unlike $(l_i, l_j)$, $l_i \rightarrow l_j$ indicates that there may be one or more locks between $l_i$ and $l_j$.

\begin{equation} \label{equ:deri2}
\begin{split}
& \forall l_i, l_j \in LS_i \text{ and } l_i \rightarrow l_j, \text{if } l_i \in LS_j \text{ and } l_j \in LS_j, \\
& \text{then } l_i \rightarrow l_j \text{ also holds in } LS_j.
\end{split}
\end{equation}

Based on the rule, we propose two ways of repair, 1) adding new locks and 2) extending the lock scope.
We first try to repair a race by adding a new lock.
If the lock order of the program violates Rule~\ref{equ:deri2}, we will try to extend the lock scope instead of adding new locks.
Otherwise, we insert a lock operation right before $L_i$ and $L_j$, and also insert an unlock operation right after $L_i$ and $L_j$.
Then we validate the lock orders of the repaired program according to Rule~\ref{equ:deri2}. 
If the new lock meets the rule, we try to fix other races.
\textit{task1} and \texttt{ISR1} in Figure \ref{fig:EPR1} show a race repaired by adding a new lock.
If Rule~\ref{equ:deri2} is violated, we try to extend the lock scope. 
In order to meet Rule~\ref{equ:deri2}, 
we first locate the closest locks after $L_i$ and $L_j$, which are denoted as $l_{ii}$ and $l_{jj}$.
Then, we extend one of the locks in $LS_{cand} = (l_{i1}, ... , l_{ii}) \cap (l_{j1}, ... , l_{jj})$, such that $e_i$ and $e_j$ can be protected by the same lock and meet Rule~\ref{equ:deri2}.
If $LS_{cand}$ is empty, we report this race pair is unable to repair by ECS.
\textit{task2} and \textit{ISR2} in Figure \ref{fig:EPR2} show a race repaired by extending a critical section.
}

\begin{figure*}[t]
\centering

\begin{subfigure}[b]{0.55\textwidth}
\centering
\begin{lstlisting}[language={[ANSI]C},  basicstyle=\ttfamily\scriptsize, numbers=left, frame=single,xrightmargin=10em]
/*           Race program 1           */
void task1(){            void ISR1(){
  spin_lock(&lock1);       
  ...                      //race code.
  spin_unlock(&lock1);     ...
  
  /* race code. */       }
  ...
  
}
\end{lstlisting}
\end{subfigure}
~~
\begin{subfigure}[b]{0.55\textwidth}
\centering
\begin{lstlisting}[language={[ANSI]C},  basicstyle=\ttfamily\scriptsize, numbers=left, frame=single,xrightmargin=10em]
/*           Repaired by AL           */
void task1(){            void ISR1(){
  spin_lock(&lock1);       spin_lock(&lock2);
  ...                      //race code.
  spin_unlock(&lock1);     ...
  spin_lock(&lock2);       spin_unlock(&lock2);
  /* race code. */       }
  ...
  spin_unlock(&lock2);
}
\end{lstlisting}
\end{subfigure}
\vspace{-5pt}
\caption{An example of adding locks. }
\label{fig:EPR1}
\end{figure*}

\begin{figure*}[t]
\centering

\begin{subfigure}[b]{0.55\textwidth}
\centering
\begin{lstlisting}[language={[ANSI]C},  basicstyle=\ttfamily\scriptsize, numbers=left, frame=single,xrightmargin=10em]
/*           Race program 2           */
void task2(){            void ISR2(){
  spin_lock(&lock1);       spin_lock(&lock2);
  ...                      ...
  spin_unlock(&lock1);     spin_unlock(&lock2);
  /* race code. */.        //race code.
  ...                      ...
  spin_lock(&lock2);     }
  ...
  spin_unlock(&lock2);
} 
\end{lstlisting}
\end{subfigure}
~~
\begin{subfigure}[b]{0.55\textwidth}
\centering
\begin{lstlisting}[language={[ANSI]C},  basicstyle=\ttfamily\scriptsize, numbers=left, frame=single,xrightmargin=10em]
/*           Repaired by ECS           */
void task2(){            void ISR2(){
  spin_lock(&lock1);       spin_lock(&lock2);
  ...                      ...
  spin_unlock(&lock1);     //race code.
  spin_lock(&lock2);       ...              
  /* race code. */         spin_unlock(&lock2);
  ...                    }
  ...
  spin_unlock(&lock2);
}
\end{lstlisting}
\end{subfigure}
\vspace{-5pt}
\caption{An example of extending critical sections.}
\label{fig:EPR2}
\end{figure*}

\subsubsection{Race repair process}\label{sec:RRP}

\begin{algorithm}[t]
\small
\caption{Race repair and validation}
\label{fig:RRV}
\begin{algorithmic}[1]
\raggedright
\REQUIRE $P$, $WNSet$
\ENSURE $P'$
\STATE locateLine($WNSet$) \label{RRV:LL}
\STATE $P'$ = $P$ /* Backup $P$ */\label{RRV:PP}
\FOR{each $WN$ $\in$ $WNSet$} \label{RRV:CRS1}
	\STATE chooseRepairStrategies($WNSet$) \label{RRV:CRS}
	\STATE $P'$ = repairProg($P'$, $WN$) \label{RRV:RP}
\ENDFOR \label{RRV:CRS2}
\WHILE{validation($P'$) is not passed} \label{RRV:Vad1}
	\STATE extendCriticalSection($P'$) \label{RRV:ECS}
\ENDWHILE \label{RRV:Vad2}
\STATE mergeFix($P'$) \label{RRV:MF}
\end{algorithmic}
\end{algorithm}

To repair a validated race, we propose a human-computer corporation based repair process.
\wymm{The overall procedure is shown in Algorithm \ref{fig:RRV}.
First, it analyzes the race report to locate its source lines (at line \ref{RRV:LL}).
After that, for each validated race warning, the algorithm chooses feasible repair strategies based on Table \ref{tab:sum} and conducts the repair (at lines \ref{RRV:CRS1} to \ref{RRV:CRS2}).
\wym{Previous works claim that the majority of the programs patched by generate-and-validate patch generation systems do not produce acceptable outputs  \cite{qi2015analysis}.}
Therefore, we propose an additional subcomponent, which validates repaired programs, in order to avoid such situations.
In this subcomponent, the algorithm validates the fixed program, either by the validation component from \Name{} or manual code review from developers (at lines \ref{RRV:Vad1} to \ref{RRV:Vad2}).
Then, the algorithm tries to extend critical sections for unfixed races (at line \ref{RRV:ECS}).
Finally, it reduces repair operations by merging critical sections (at line \ref{RRV:MF}).
Next, we discuss how to validate repaired programs and reduce repair operations.}

\textbf{Repaired program validation:} 
\wymm{To validate repaired programs, we reuse the first three components (\ie, static analysis, guided symbolic execution and dynamic validation) of \Name{}. }
Each generated patch is validated by \Name{} to check whether the bug has been fixed. 
Specifically, we check whether the dynamic validation component still reports the race.

Note that patches generated by \Name{} are still required to be validated.
The reason for validation is that the position of inserting locks (or interrupt controlling operations) are lines nearest the racing points. 
\wymm{Although it can repair the races defined in section \ref{sec:Def}, it is not enough to repair atomicity violations, which are the combination of two or more races of the same shared resource.
For example, there are two races and one atomicity violation in the code \texttt{Task()\{write(sv);read(sv);\} ISR()\{write(sv);\}}. 
Adding locks to create separate critical sections can fix the races but cannot fix the atomicity violation.
Therefore, after repairing, we generate a patched program and validate it by the dynamic validation phase. 
If we can still detect races or failures, we will gradually enlarge the corresponding critical sections according to the repair algorithms until no bugs and no failures.
In this way, extending critical sections could fix the atomicity violation in the example. 
The repaired program is \texttt{Task()\{lock(l);write(sv);read(sv);unlock(l);\} ISR()\{lock(l);write(sv);unlock(l);\}}.
}

\textbf{Reduce repair operations:} After repairing all races, we analyze critical sections created by \texttt{IDE}, \texttt{AL} or \texttt{ECS}. 
Then, we merge two critical sections if one of the following conditions are met: 
1) the two critical sections have overlap.
2) there is no instruction between two critical sections.
We only merge critical sections with the same IRQ if they are created by \texttt{IDE}.
When merging critical sections created by \texttt{AL} or \texttt{ECS}, we replace two locks with the same lock.
Merging critical sections reduces the number of repair operations by removing redundant operations.

\subsection{Implementation}
%
The static analysis component of \Name{} was implemented using
the Clang Tool 3.4 \cite{Clangtool}.
Our alias analysis leveraged the algorithm in  \cite{andersen1994program}
to handle the alias of shared resources.
Our guided symbolic execution was implemented based on KLEE 1.2~\cite{KLEE} 
with STP solver~\cite{STP} and KLEE-uClibc~\cite{KLEEuclibc}.
Since most kernel functions are not supported by KLEE and KLEE-uClibc,
we have extended KLEE-uClibc to support kernel functions such as \emph{request\_irq()}.
In order to guide the symbolic execution toward specific
targets (\ie, potential racing points), we modified KLEE to only gather
constraints related to the paths that are generated by static analysis.
\wymm{We used Simics virtual platforms with a simulated X86 CPU to implement the dynamic validation phase. }
Simics provides APIs that can be accessed via Python scripts 
to monitor concurrency events and to manipulate memory 
and buses directly to force interrupts to occur. 
\wymm{Finally, our automatic race repair component was implemented in Python to repair races from the dynamic validation phase.}

\section{Empirical Study}
\label{sec:study}

To evaluate \Name{} we consider three research questions:

\vspace*{3pt}
\noindent
{\bf RQ1:}
How effective is {\sc \Name{}} at detecting interrupt-level
race conditions?

\vspace*{3pt}
\noindent
{\bf RQ2:}
How efficient is {\sc \Name{}} at detecting interrupt-level
race conditions?

\vspace*{3pt}
\noindent
{\bf RQ3:}
How effective and efficient 
is {\sc \Name{}} at repairing interrupt-level
race conditions?

\vspace*{3pt}

\noindent
RQ1 allows us to evaluate the
effectiveness of  our approach in terms of
the number of races detected at different phases,
and their abilities to reduce false positives.
RQ2 lets us consider the efficiency of our approach
in terms of analysis/testing time and platform overhead.
RQ3 examines whether our repair strategy is 
effective and efficient.

\subsection{Objects of Analysis}

As objects of analysis, we chose both open source
projects and industrial products. 
First, we selected 118 device driver programs that can be compiled
into LLVM bitcode from 
four versions of Linux Kernel.  
We next eliminated from consideration those drivers that 
could not execute in Simics environment; this process left
us with four drivers: keyboard, mpu401\_uart, i2c-pca-isa, and 
mv643\_eth. The 114 drivers were not executable because their
corresponding device models were not available in Simics -- they need
to be provided by developers.
As part of the future work, we will develop new device models for Simics in order to study more device driver programs. 

\ignore{Additionally, the number of subject is the intersection of available drivers in host OS (used for Static Analysis and Guided Symbolic Execution) and guest OS (used for Dynamic Validation).
}

We also selected two driver programs from LDD \cite{corbet2005linux}: short and 
shortprint. To create more subjects, we  manually seeded a concurrency
fault to each of the two LDD programs. Specifically,
we injected a shared variable increment operation and a decrement 
operation in their interrupt handlers.
The fault injection did not change the semantics of the original 
programs but induced new races to these programs.
The two programs are denoted as 
short (EI) and shortprint (EI).

The other three subjects
are real embedded software from  China Academy of Space Technology. 
Module1 is an UART device driver. 
Module2 is a driver for the lower computer.
Module3 is used to control the power of engine.
Table~\ref{tableRes} lists all eleven programs,
the number of lines of non-comment code they contain,
the number of interrupts (with different
priorities), the number of functions, the number of
shared resources, and the number of basic blocks.
The number of basic blocks indicates the 
complexity of symbolic execution.
The size of the benchmarks is consistent with a
prior study of concurrency bugs in device driver programs~\cite{vojdani2016static},
which ranges from less than a hundred line of code to 
thousands of lines of code.

All our experiments were performed on a PC with 4-core Intel Core
CPU i5-2400 (3.10GHz) and 8GB RAM on Ubuntu Linux 12.04.
For the simulation, the Host OS was Ubuntu 12.04 and the guest OS was 10.04.
Simulation was based on real-time mode and conducted without VMP 
(In order to run Intel Architecture (IA) targets quickly on IA-based hosts.). 
The timeout for symbolic execution was set to 10 minutes. 

\begin{table}[!t]
\centering
\small
\caption{Objects of Analysis}
\begin{tabular}{|l|r|r|r|r|r|}
\hline
Program name        & \emph{LOC}   & \emph{\#INT}   & \emph{\# Func}   &
\emph{\#SR} & \emph{\#BB} \\\hline 
keyboard\_ driver     & 84    & 1     & 4     & 5     & 45        \\\hline %
mpu401\_ uart   & 630   & 1     & 16    & 2     & 316       \\\hline 
i2c-pca-isa    & 225   & 1     & 11    & 9     & 111      \\\hline 
mv643xx \_eth.c       & 3256  & 1     & 29    & 7    & 1076      \\\hline 
short               & 704   & 5     & 18    & 20    & 315       \\\hline %
shortprint          & 531   & 1     & 11    & 22    & 266       \\\hline %
short (EI)          & 707   & 5     & 18    & 20    & 317     \\\hline %
shortprint (EI)     & 530   & 1     & 11    & 22    & 266      \\\hline %
module1             & 168   & 1     & 3     & 1     & 55      \\\hline 
module2             & 154   & 2     & 7    & 4     & 62        \\\hline 
module3             & 99   & 2     & 8     & 1     & 40      \\\hline 
\end{tabular}
\label{tableRes}
\end{table}%

\subsection{Dependent Variables}
  
We consider several measures 
(\ie, dependent variables) to answer
our research questions.
Our first dependent variable measures
technique \emph{effectiveness} in terms of the {\em
number of races detected}.  
We measure the number of races detected
in each of the three phases. 
\wymm{Similar to data races in multi-thread programs, the race condition we detect also has benign races and harmful races.}
We 
also inspected all of the reported real races
(from the dynamic validation phase)
that did not result in detectable failures to determine
whether they were harmful or benign. 

To assess the {\em efficiency} of techniques
we rely on four dependent variables, each 
of which measures one facet of efficiency.
The first dependent variable measures 
the analysis and testing time 
required by \Name{} across the three phases. 
Although measuring time is undesirable in 
cases in which there are nondeterministic 
shared resource accesses among processes,
this is not a problem in our case because we use 
a VM that behaves in a deterministic manner. 

Our second variable regarding efficiency
measures the extra
{\em platform overhead} associated with \Name{}.
This is important because using virtual platforms 
such as Simics for testing can increase costs, 
since virtualization times can be longer 
than execution times on real systems.
We calculate platform overhead by dividing 
the average runtime per test run on Simics 
by the runtime per test on the real machine. 
Note that judging whether races are harmful is not 
taken into account when computing the overhead of 
\Name{} because it is independent of 
techniques for locating harmful races.

To measure the effectiveness of repair,
we patch the program as suggested by \Name{} 
and run its test cases (generated from
symbolic execution). We consider a repair is
valid if it does not fail any test case in
the dynamic validation.
To measure the efficiency of repair, 
we compare the program execution time
without the patch to the execution time with
the patched applied. 

\subsection{Threats to Validity}
The primary threat to external validity for this study involves 
the representativeness of our programs and faults.  
Other programs may exhibit different behaviors and 
cost-benefit tradeoffs, as may other forms of test suites.  
However, the programs we investigate are 
widely used and the races we consider are real
(except the seeded races on the two LDD programs). 

The primary threat to internal validity for this study
is possible faults in the implementation of our approach
and in the tools that we use to perform evaluation.
We controlled for this threat by extensively testing
our tools and verifying their results against smaller
programs for which we can manually determine the correct results.
We also chose to use popular and established tools (\eg, Simics
and KLEE) to implement the various modules in our approach.
%
As an additional threat to internal validity, race
manifestation can be influenced by the underlying
hardware~\cite{Osterman05,arch1}. For example,
microprocessors that provide virtualization support may
be able to prevent certain races from occurring due to
fewer system calls. Our work uses \textsc{Simics}, a
full platform simulator to provide us with the
necessary controllability and observability to cause
races.  Simics has been widely used to expose
difficult-to-reproduce faults including
races~\cite{Windriver}.  The version of \textsc{Simics}
that we used does not simulate the later Intel
processors with hardware virtualization support---a
feature that can affect our ability to produce races.
Nonetheless, our system was able to detect
previously documented races existing in our
experimental subjects.  Therefore, the execution
patterns seen using \textsc{Simics} should be
comparable to those that would be observed in the real
systems. 

Where construct validity is concerned, numbers 
of races detected are just one variable 
of interest where effectiveness is concerned. 
Other metrics such as the cost of manual analysis could be valuable.
\wym{Furthermore, the performance also depends on the experiment setup, such as the time of unrolling loops, symbolic execution timeout, maximum repair attempts, \etc
}

\section{Results and Analysis}
\label{sec:result}

Table~\ref{efficResult} reports the effectiveness and efficiency
results observed in our study; 
we use this table to address our research questions.


\begin{table*}[!t]
\centering
\small
\caption{Experimental Results}
\begin{tabular}{|r|r|r|r|r|r|r|r|r|r|}
\hline
\multirow{2}{*}{Programs} & 
\multicolumn{4}{c|}{Race Detected}  & 
\multicolumn{3}{c|}{Execution Time (second)}  & 
\multirow{2}{*}{\tabincell{c}{Sim\\Overhead}} & 
\multirow{2}{*}{\tabincell{c}{Dyna\\Only}} \\
\cline{2-8}
& \tabincell{c}{Static\\Analysis} & 
\tabincell{c}{Symbolic\\Execution} & \tabincell{c}{Dynamic\\Validation} 
& \tabincell{c}{Manual\\Checking} & \tabincell{c}{Static\\Analysis} & 
\tabincell{c}{Symbolic\\Execution} & \tabincell{c}{Dynamic\\Validation}  
&  &\\\hline
keyboard\_driver  & 4 & 4& 4& 4   & 0.073   & 1.03     & 1.65  & 892x 
& 4 \\\hline 
mpu401\_ uart  & 146 & 129 & 47 & 47 & 0.088   & 1251.83      & 75.2  & 245.3x
&12   \\\hline 
i2c-pca-isa    & 4 & 4& 1& 1     & 0.078   & 1.00     & 42.1  & 530.1x 
&1   \\\hline 
mv643xx \_eth & 16 & 14& 10& 10     & 0.183   & 1207.97     & 102.2 & 64.4x  
&2 \\\hline 
short        & 127 & 35& 18& 18       & 0.109   & 41.53     & 26.8  & 297x 
&14  \\\hline %
shortprint   & 4 & 2& 0& 0       & 0.088   & 1.25     & 21.61 & 445.6x 
&0 \\\hline %
short (EI)   & 149 & 41& 24& 24       & 0.106   & 48.28     & 24.13 & 285.6x 
&18   \\\hline %
shortprint (EI) & 14 & 8& 6& 6    & 0.091   & 4.44     & 49.3  & 425.8x 
&6  \\\hline %
module1     & 4 & 4& 4& 4        & 0.076   & 0.91     & 1.54  & 669.2x 
&4   \\\hline 
module2     & 93 & 65& 64& 64        & 0.075   & 21.48     & 1.25  & 590.1x 
& 64  \\\hline %
module3     & 15 & 15& 12& 12        & 0.073   & 3.39     & 1.06  & 426x 
& 12   \\\hline 

\end{tabular}
\label{efficResult}
\end{table*}%

\begin{table}[!t]
\centering
\small
\caption{Repair Results}
\begin{tabular}{|r|r|r|r|r|}
\hline
\multirow{2}{*}{Programs} & 
\multicolumn{2}{c|}{\#New operations} &
\multicolumn{2}{c|}{Repair Overhead} \\
\cline{2-5}
& \tabincell{c}{IDE} &  \tabincell{c}{AL/ECS} & \tabincell{c}{IDE} &  \tabincell{c}{AL/ECS} \\\hline
keyboard\_driver & 2 & 4  & 0x & 0x \\\hline 
mpu401\_uart  & 28 & 46 &  0.04x & 0.01x   \\\hline 
i2c-pca-isa    & 2 & 4 & 0.75x & 0.01x   \\\hline 
mv643xx\_eth   & 20 & 26 & 0.01x & 0x \\\hline 
short          & 14 & 28 & 0.03x & 0.01x  \\\hline %
shortprint     & NA & NA & NA & NA \\\hline %
short (EI)     & 16 & 30 & 0.03x & 0.01x   \\\hline 
shortprint (EI)& 6 & 10 & 0.05x & 0.01x  \\\hline %
module1        & 2 & 4 & 0.8x  & 0.01x   \\\hline 
module2        & 12 & 18 & 0.07x & 0.01x  \\\hline %
module3        & 4 & 10 & 0.08x & 0.01x   \\\hline 

\end{tabular}
\label{tab:rep}
\end{table}%

\subsection{RQ1: Effectiveness of \Name{}}

Columns~2-5 in Table~\ref{efficResult} show
the number of races reported by static
analysis, the number of races remained after
symbolic execution,
the number of real races reported by the dynamic validation across
all 11 subjects, and the number of true races validated manually by us. 
\wymm{We reported the detected races in the 9 subjects to developers
and 4 of them were confirmed (\ie, keyboard\_driver, module1, module2 and module3). 
Others are waiting for the
confirmation. 
Note that we do not report bugs in the 2 subjects with injected 
races because they are introduced to evaluate the effectiveness of \Name{}.}

\wym{
\wymm{
As the results show, 
the symbolic execution reduced the number of false
positives contained in the sets of static race warnings by 40.3\% overall,
with reductions ranging from 0\% to 96.6\%
across all 11 subjects. The dynamic validation reduced the number of
races reported by symbolic execution by 36.7\% overall,
with reductions ranging from 0\% to 100\%. }
The manual examination revealed that
among all real races (\ie, exclude inject bugs) reported by the dynamic validation,
all races are real and harmful.}
In total, \Name{} detected 190 races. Only on {\tt shortprint}
did \Name{} not detect any races. No false negatives were found
on all programs by the manual inspection. 
 
On two out the 11 subjects, symbolic execution reported
equal number races to the dynamic validation
({\tt keyboard\_driver} and {\tt module1}).
In other words, symbolic execution did not report
false positives on the two programs.
On the other nine programs, symbolic execution 
did report false positives. By further examining the programs,
we found two reasons that led to the false positives.
The first reason is due to the unknown access type
(read and write) in external
functions. For example, on {\tt mpu401\_uart}, 
the ISR calls an external library ({\tt snd\_mpu401\_input\_avail}) 
taking an SV as the argument. The symbolic execution 
treats this access as a write since 
static analysis incorrectly identifies it as a write. 
The second reason is due to the conflict path constraints
between the main task and ISRs, which 
resulted in time-out. 
In this case, the race reported by the static
analysis is directly sent to the dynamic 
validation phase.
The third reason is that it is incapable of
recognizing the implicit interrupt operations;
This case happened to the program {\tt short} . 

\subsection{RQ2: Efficiency of \Name{}}

Columns 6-8 in Table~\ref{efficResult} report the 
analysis time of
static analysis, symbolic execution, and dynamic validation.
On two programs ({\tt mpu401\_uart} and  {\tt mv643xx\_eth}), 
the symbolic execution reached the time
limit (\ie, 10 minutes) on the two static warnings
of each program due to the unsolvable path constraints.
Therefore, their times of symbolic
execution were much higher than the other programs.
Overall, the total testing time spent by \Name{}
ranged from 2 seconds to 23 minutes across all 11
subjects. Specifically, the time for static 
analysis never exceeded
0.2 second, which accounted for less than 0.01\% of
total testing time overall. The time spent on 
symbolic execution was 235 seconds in arithmetic mean, accounting
for 88.1\% of total testing time.
The remaining (31 seconds) time was spent on dynamic validation,
which accounted for 11.8\% of total testing time. 
The time for symbolic execution and
dynamic validation varied with the number 
of detected static warnings.

\Name{} incurred platform overhead due to the use of VMs.
Column~9 of Table~\ref{efficResult} lists
the average platform overhead associated with
\Name{} across all test runs.
As the table shows, the average platform 
overhead ranged from 64x to 669x.
As we can see from the result, the less complex a subject is,
the more platform overhead it incurred. This is because
our execution observer was implemented using
the callback functions provided by the Simics VM;
it took time for the MV to trigger callback functions.
However, considering the benefits of virtual platforms
and the difficulty of detecting interrupt-level race conditions,
such overhead is trivial.



\subsection{RQ3: Effectiveness and Efficiency of Automated Repair}

We patched the program as suggested by \Name{} 
and run its test inputs along with the
controlled interrupts. We consider a repair is
valid if it does not fail any test case 
(\ie, no races are reported by  \Name{}).
Our results indicate that the repairs on all
eleven programs are valid and did not 
incur new races or deadlock.
\wym{To measure the repair overhead, we randomly choose three inputs from guided symbolic execution (including interrupt trigger time) and average the performance overhead.} 

\wymm{Columns 2-3 of Table~\ref{tab:rep} show the number of new operations inserted to fix races. 
The repair for \textit{shortprint} is not appliable since there is no report.
The number of new operations is not very high even for programs with many races.
The main reason is that we merge critical sections to reduce redundant interrupt or lock operations.
For example, there are two races in \textit{task()\{read(x); read(y);\}} and \textit{ISR()\{write(x); write(y);\}}, which can be fixed by \textit{task()\{lock(mut); read(x); read(y); unlock(mut);\}} and \textit{ISR()\{lock(mut); write(x); write(y); unlock(mut);\}}.
Note that there are only 4 new operations instead of 8 operations.
}
Columns 4-5 of Table~\ref{tab:rep} reports the overhead of the repair.
On nine out of 11 programs, the overhead was
less than 0.09.  These overheads are in the same order 
of magnitude as that of other thread-level
concurrency fault repair techniques~\cite{jin2011automated,liu2012axis}. 
We consider such small overheads to be negligible. 
On the other hand, the overheads
on the {\tt module1} and the {\tt i2c-pca-isa} are much higher. 
The reason is that after enforcing the interrupt disable and enable
operations, it changed the control flow of the main tasks since it was preempted by ISR, causing
the main tasks to execute longer. 

Note that different interrupt request lines induced different overhead.
In the experiment, we found that the higher priority interrupt 
tends to induce higher overhead when disabling it.
We conjecture that disabling higher priority interrupts
are likely to prevent lower priority ISRs from being
executed and thus cause interrupt latencies. 

Finally, we report fixed versions of {\tt module1}, {\tt module2} and {\tt module3} to the developer and all of them are confirmed.

\ignore{
We instrument interrupt enable and disable operations to verify the effectiveness
of automatic repair. After automatic race repair,
none of validated data races occur even we forcible trigger corresponding interrupts by simulation.

We implemented two ways to simulate embedded programs: ``out-of-tree''
build and ``in-tree'' build. ``out-of-tree'' build means that the source
for the kernel module is somewhere in our home directory structure,
not the kernel source directory like the standard device drivers that
come with Linux. The other difference between ``out-of-tree'' build
and ``in-tree'' build is that ``in-tree'' build requires less
human intervention, but we need to compile the kernel rather
than just compile kernel module.
}

\section{Discussion}
\label{sec:dis}

In this section, we first summarize our 
experimental results and then explore additional 
observations and limitations relevant to our study. 


\subsection{Summary of Results}
\Name{}'s static analysis component
can detect potential race conditions with
a false positive rate 72.0\%. Our static analysis is able to 
handle nested interrupts with different priorities, as opposed to 
deal with race conditions only between tasks and ISRs~\cite{chen2011static}.
\Name{}'s symbolic execution reduced the false positive rate 
to 49.8\%. 
The VM-based dynamic validation eliminated all false
positives. 
\wym{Meanwhile, we manually validated the results from \Name{} and found that they are all true races.
The average testing time is 4.5 minutes on each program. }
\Name{}'s
automated race repair component  effectively
repaired all detected race conditions while inducing little overhead. 
There are only two exceptions, which are i2c-pca-isa and module1. The reason is that the functions contain race pair is too simple to ignore the overhead of introducing additional interrupt controlling operations. 
Therefore, adding locks is a better choice but adding interrupt controlling operations is simpler.

If these results generalize to other real objects, then {\em if engineers wish to target 
and repair race detection  in interrupt-driven
embedded system, \Name{} is a cost-effective technique to
utilize.} In the case of non-existing VMs, developers
can still use static analysis and symbolic execution
to detect races. 

\ignore{In \Name{},
static analysis, symbolic execution and dynamic validation 
are supplementary to each other. 
For example, we do not make any instrumentation when
using ``in-tree'' build. Therefore, we can not ensure the execution
path is expected. However, if static analysis indicates that shared
resources only access once in tasks and interrupt handlers.
We can ensure the execution path is the same as the path
indicated by static analysis.
}

\subsection{Further Discussion} 

\subB{Influence of test input generation.}
There have been techniques for detecting 
concurrency faults that occur due to interactions between application and 
interrupt handlers~\cite{Regehr05, yu2012simtester, lai2008inter,
AST10Higashi}. However, these techniques neither handle
nested interrupts nor considers priority constraints
among tasks and ISRs. Also, they
do not have the static analysis and symbolic
execution components, which could miss races
that can only be revealed by certain inputs. 
In addition, these techniques are not applicable
in the case of non-existing VMs or runtime
environment. To further investigate whether the use of static analysis
and symbolic execution can improve the race detection
effectiveness, we disabled the two components and did 
see missing races. 
Columns~10 in Table~\ref{efficResult}
reports the numbers of races detected when
using only the dynamic validation component.
\wym{When only the dynamic validation stage is used, we regard it as a dynamic testing tool since it can detect races. We regarded the test subjects as a black box and manually fed inputs.  }
As the data shows, in total, it
detected only 137 races -- 28.2\% less effective than
\Name{}.

\subB{Atomicity violations.}
\Name{} considers one type of definition of race conditions -- order violations.
In practice, testers can adopt different definitions because there is not a single general definition for the class of race conditions that occur between an $ISR$ and a task/an $ISR$.
\Name{} may miss faults due to atomicity violations.
For example, if a read-write shared variable pair in the main 
program is supposed to be atomic, the $ISR$ can read 
this shared variable before it is updated in the main program. 
Since \Name{} does not capture the read-read access pattern,
this fault may be missed.

\ignore{In the experiment, 
we found that a race condition related to the atomicity 
violation was missing. Figure \ref{falsenage} shows this example.
The code region in the {\tt put} function
is supposed to execute atomically.
However, the execution can be preempted by function \textit{isr} right
before line 4. As results, the 
the variable {\tt StepsTemp}
in the ISR reads the wrong value {\tt 0xffff}, which is supposed
to be {\tt 0}.
}


\ignore{
\lstset{
  language={[ANSI]C},
  moredelim=**[is][\color{OliveGreen}]{@}{@},
  basicstyle=\fontsize{6}{6}\ttfamily, mathescape,
  breaklines=true,
  showstringspaces=false
  }
\noindent
\begin{figure}[t]
\footnotesize
\begin{mdframed}[roundcorner=5pt]
\scriptsize
\begin{lstlisting}[language=C++,showstringspaces=false]
void put()
{
  if(DistanceSteps == 0xffff)
    DistanceSteps=0; 
}
void isr()
{
  StepsTemp=MAXSTEP-DistanceSteps; 
}

\end{lstlisting}

\vspace*{-4pt}
\end{mdframed}
\vspace*{-5pt}
\caption{\label{falsenage} \textbf{\small False negative in dynamic validation}}
\normalsize
\vspace*{-5ex}
\end{figure}
}

\subB{Inline functions.}
In the dynamic validation phase, we use memory breakpoints
to detect when concurrency events are executed.
However, some simple functions are optimized as inline functions
by compilers. In this case, breakpoints for these functions 
cannot be triggered. To handle this case, we need to disable 
optimization for these functions.

\subB{Dynamic priority assignment.}
Many false positives in the 
static analysis phase are caused by nested interrupts,
because \Name{} does not recognize priorities
that are dynamic assigned. 
These false positives can result in more validation
time in symbolic execution and dynamic simulation.
As part of future work, we will consider operations
involving dynamic priority adjustment.

\subB{Complex repair strategies.}
As we can see from the survey of race repairs, 
some repair strategies (\eg, changing operation orders, 
adding additional checks) depend on semantics
of a program, which is difficult to propose an accurate repair method
to support these strategies.
However, among the top 4 repair strategies, the two semantics-based repair strategies (\ie, COO and AAC) strategies take 50.4\% among all repair methods, 
which means that both of them are important and popular repair methods. 

\subB{Scalability to the entire system.}
%
In our study, the analysis involves a test
program, the interrupt handler that interacts with the
device driver, and the device driver code.  The
key point here is that the tester focuses on a specific
component~\footnote{A component is a device driver program. The list of components can be identified by popular Linux commands such as ``modprobe"} and how it interacts with the rest of the
components.  If the focus changes to a different
component, the same analysis can be applied to test the
new component. As such, the proposed approach is more suitable for component testing
instead of testing the entire system at once. 

\subB{Multi-core systems.}
\wymm{
Our experiment focuses on single-core systems.
When it comes to multi-core systems,  enabling/disabling interrupts is on a per-core basis \cite{ganguly1999theory}.
Therefore, it is difficult to globally disable interrupts on all cores.
In this case, our method can detect races happening across cores because the way of detecting shared resources in the dynamic validation phase is to monitor the memory address of shared resources.
As for repairing races in multi-core systems, disabling interrupts is not a choice but adding locks or extending critical sections can fix this kind of race.
}


\section{Related work}
\label{sec:ret}

There has been a great deal of work on analyzing,
detecting, and testing for thread-level data
races~\cite{STVR15,Maiya14,Raman12,
Sen08,
yu2005racetrack,savage1997eraser,o2003hybrid,pozniansky2003efficient,
duesterwald1991concurrency, callahan1990analysis,
aspinall2007formalising, manson2005java}. 
However, as discussed in Section~\ref{subsec:thread},
existing techniques on testing
for thread-level concurrency faults have rarely
been adapted to work in scenarios
in which concurrency faults occur due to asynchronous
interrupts. 
\ignore{It is unclear whether these approaches can
work in such a scenario for two reasons.  First,
controlling interrupts requires \emph{fine-grained}
execution control; that is, it must be possible to
control execution at the machine code level rather than
the program statement level, which is the granularity
at which many existing techniques
operate~\cite{Yu12maple,Sen08}. Second, occurrences
of interrupts are highly dependent on hardware states;
that is, interrupts can occur only when hardware
components are in certain states. Existing techniques
are often not cognizant of hardware
states~\cite{AST10Higashi, Regehr05}.
}

\subB{Dynamic Testing in interrupt-driven programs.}
There are several techniques for testing embedded
systems with a particular focus on interrupt-level 
concurrency faults~\cite{Regehr05, yu2012simtester, lai2008inter,
AST10Higashi}. For example, Regehr et
al.~\cite{Regehr05} use random testing to test Tiny OS applications.  
They propose a technique called restricted interrupt discipline (RID)
to improve naive random testing (\ie, firing
interrupts at random times) by eliminating aberrant
interrupts. However, this technique
is not cognizant of hardware states and may lead to 
erroneous interrupts.  SimTester \cite{yu2012simtester} leverages VM to address
this problem by firing interrupts conditionally 
instead of randomly. Their
evaluation shows that conditionally fired
interrupts increase the chances of reducing cost.  
However, all the foregoing techniques do not
consider interrupt-specific event constraints
(\eg, priorities) and may lead to imprecise results.
In addition, they are incapable of
automatically generating test inputs or repairing race conditions.
In contrast, our approach can cover all
feasible shared variables in the application instead of
using arbitrary inputs; this can help the program
execute code regions that are more race-prone.


\subB{Static Analysis in interrupt-driven programs.}
There has been some work on using static analysis to 
verify the correctness of interrupt-driven 
programs~\cite{wei2011static, chen2011static, regehr2007interrupt, fmcad11}.
For example, Regehr et al. \cite{regehr2007interrupt} propose a method to 
statically verify interrupt-driven programs. Their work first 
outlines the significant ways in which interrupts 
are different from threads from the point of view of verifying the absence of race conditions.
It then develops a source-to-source transformation method to
transform an interrupt-driven program into a semantically 
equivalent thread-based program so that a thread-level static race
detection tool can be used to find race conditions, which is
the main benefit of their approach. Comparing to \cite{regehr2007interrupt}, \Name{} has two 
advantages.  First, proof of the correctness of code transformation is often non-trivial; 
\cite{regehr2007interrupt} does not provide  proofs showing the transformation is correct
or scalable. In contrast, \Name{} is transparent and does not require 
any source code transformation or instrumentation and can be applied
to the original source code. Second,  \Name{}  uses dynamic analysis 
to validate warnings reported by static race detectors. 
Our evaluation showed that \Name{} can eliminate a large portion
of false positives produced by static analysis, 
whereas Regehr's work \cite{regehr2007interrupt}  on 
seven Tiny OS applications does not evaluate the precision of
their technique.

Jonathan et al.~\cite{fmcad11} first statically 
translate interrupt-driven programs into sequential 
programs by bounding the number of interrupts, 
and  then use testing to measure execution time. 
While static analysis is powerful, it can
report false positives due to imprecise local
information and infeasible paths. In addition, as embedded
systems are highly dependent on hardware, it is
difficult for static analysis to annotate all
operations on manipulated hardware bits; moreover,
hardware events such as interrupts usually rely on
several operations among different hardware bits.
\Name{} leverages the advantages of static analysis
to guide precise race detection. 
Techniques combined with static and dynamic method \cite{wang2015detecting}
could also detect and verify races. However, due to the lack of test case generation
method, Manually efforts are required to inspect codes and generate test cases
to reach race points.

\subB{Dynamic Testing in event driven programs.}
There has been some research on testing for concurrency faults in event-driven
programs, such as mobile applications~\cite{Hsiao2014, Maiya14, Hu2016, bielik2015scalable} 
and web applications~\cite{raychev2013effective,hong2014detecting}. 
Although the event execution models of event-driven and interrupt-driven
have similarities, they are different in several ways. 
First, unlike event-driven programs that maintain an event queue as first-in,
first-out (FIFO) basis, interrupt handlers are often assigned to different priorities 
and can be preempted. Second, interrupts and their priorities
can be created and changed dynamically and such dynamic behaviors
can only be observed at the hardware level. 
Third, the events in event-driven programs are employed
at a higher-level (\eg, code), whereas hardware interrupts happen at a 
lower-level (\eg, CPU); interrupts can occur only when hardware
components are in certain states. The unique characteristics of interrupts 
render inapplicable the existing race detection techniques for event-driven
programs. 


\subB{Hybrid techniques.}
There has been some research on 
combining static analysis and symbolic execution
to test and verify concurrent programs~\cite{Farzan13,sen2006cute,
samak2015synthesizing, samak2016directed, Guo15, wang2017automatic}.
For example,  Samak et. al. \cite{samak2016directed}  
combine  static and dynamic analysis to synthesize 
concurrent executions to expose concurrency bugs.
Their approach first employs static analysis to identify the intermediate goals 
towards failing an assertion and then uses 
symbolic constraints extracted from the 
execution trace to generate new executions that satisfy pending goals.
Guo et al.~\cite{Guo15} use static analysis to identify program paths that do
not lead to any failure and prune them away during symbolic execution.
However, these techniques focus on multi-threaded
programs while ignoring concurrency faults
that occur at the interrupt level. 
As discussed in Section~\ref{subsec:thread}, interrupts are different from
threads in many ways.  On the other hand, we can 
guide \Name{} to systematically explore interrupt
interleavings or to target failing assertions.

 \ignore{
There has been several techniques on test generation~\cite{samak2015synthesizing, samak2016directed}.
Samak et. al. \cite{samak2015synthesizing, samak2016directed}
share the idea to
learn from sequential executions which concurrent interactions
to explore. In other words, they proposed sequential test-based approaches that 
execute existing sequential unit tests of
the thread-safe class. In this way they can identify concurrency bugs that may occur
when combining multiple sequential tests into concurrent tests,
and then synthesize such tests.
Our work differs from these approaches by directly guide
symbolic execution to paths indicated by static warnings.
\Name can also explore simple thread interleavings.
In contrast, \cite{samak2015synthesizing} does not consider path conditions and 
thread interleavings to expose failing executions, which result in false positives. 
Although \cite{samak2016directed} took path conditions and thread interleavings into consideration,
it selected targets (assertions going to be violated) without guidance cannot guarantee complexity/efficiency of the process.
Additionally, assertion is not enough for data race in device drivers, which requires other methods (\eg, static analysis) to locate races.
Finally, iterations of execution and constraint solving impose a relatively high computational cost.
}

\subB{Automatic race repair techniques.}
There have been several techniques on automatically repairing concurrency
faults in multi-threaded applications
~\cite{kelk2013automatically, jin2011automated, liu2012axis, liu2016understanding, surendran2014test}.
For example, AFix \cite{jin2011automated} can fix
single-variable atomicity violations by
first detecting atomicity violations, and then 
construct patches by adding locks. 

CFix \cite{jin2012automated} is designed to have more general repair
capabilities than AFix. CFix addresses the problem of fixing concurrency bugs by adding locks or condition variables to synchronize program actions.
HFix \cite{liu2016understanding} is proposed to address the problem that fixing concurrency bugs by adding lock operations and condition variables increase the fix complexity in some cases. 
HFix produces simpler fixes by reusing the code that is already present in the program rather than generating new code.
ARC \cite{kelk2013automatically}  employs a genetic algorithm 
without crossover (GAC) to evolve variants of an incorrect 
concurrent Java program into a variant that fixes all known bugs.
However, none of existing techniques have been proposed to 
repair concurrency faults in interrupt-driven embedded programs.
Surendran et al. \cite{surendran2014test} present a technique that focuses on
repairing data races in structured parallel programs.
This technique identifies
where to insert additional synchronization statements that can prevent
the discovered data races.

Interrupt-driven programs require dedicated repair strategies (\eg, control interrupts), in order to achieve better performance.
Moreover, instead of only applying repair strategies for race instructions, 
\Name{} also validates fixed programs to ensure that races are correctly addressed.
Although some techniques (\eg, AFix \cite{jin2011automated}) also 
verify their patches, the effectiveness depends on the fault detection 
tools while the detection and validation components in \Name{} 
further improve the effectiveness.

\section{Conclusion and further work}
\label{sec:clu}
This paper presents \Name{}, an automated tool to detect, validate and repair race conditions in interrupt-driven embedded software. \Name{} first employs static analysis to compute static race warnings. It then uses a guided symbolic execution to generate test inputs for exercising these warnings and eliminating a portion of false races. Then, \Name{} leverages the ability of virtual platforms and employs a dynamic simulation approach to validate the remaining potential races. Finally, \Name{} automatically repairs the detected races.
We have evaluated \Name{} on nice real-world embedded programs and showed that it precisely and efficiently detected both known and unknown races. Therefore, it is a useful addition to the developers' toolbox for
testing for race conditions in  interrupt-driven programs. 
It also successfully repaired the detected races with little performance overhead. In the future, we will further improve the accuracy of static analysis. We also intend to extend our approach to handle other types of concurrency faults.



\bibliographystyle{abbrv}
\begin{flushleft}
\bibliography{sigproc,tingting}
\end{flushleft}

\balance

\vfill

\end{document}